\documentclass[twocolumn,showpacs,preprintnumbers,amsmath,amssymb, superscriptaddress]{revtex4-1}

\usepackage{amsmath}    
\usepackage{amssymb}
\usepackage{graphicx}   
\usepackage{verbatim}   
\usepackage{color}      
\usepackage{subfigure}  
\usepackage{hyperref}   
\usepackage{blindtext}
\usepackage[normalem]{ulem}
\usepackage{natbib}

\hypersetup{colorlinks,linkcolor=blue,urlcolor=blue,citecolor=blue}
\newcommand{\beq}{\begin{equation}}
\newcommand{\eeq}{\end{equation}}
\newcommand{\bea}{\begin{eqnarray}}
\newcommand{\eea}{\end{eqnarray}}
\newcommand{\blabel}{\,b}
\newcommand{\alabel}{\,a}

\newcommand{\e}{\text e}

\newcommand{\hc}{\text{h.c.}}

\newcommand{\xs}{X_{12}^\sigma}
\newcommand{\xt}{X_{12}^{\tau}}

\newcommand{\dfi}{\text d\varphi}
\newcommand{\dd}{\text d}
\newcommand{\tr}{\tilde r}

\newcommand{\meV}{\,{\rm meV}}

\newcommand{\bK}{\mathbf K}
\newcommand{\bk}{\mathbf k}

\newcommand{\br}{\mathbf r}

\begin{document}

\title{Wigner crystal phases in confined carbon nanotubes}
\author{L. S\'ark\'any}
\affiliation{BME-MTA Exotic Quantum Phases Research Group, Institute of Physics, Budapest University of Technology and Economics, 
Budafoki \'ut 8., H-1111 Budapest, Hungary}
\author{E. Szirmai}
\affiliation{BME-MTA Exotic Quantum Phases Research Group, Institute of Physics, Budapest University of Technology and Economics, 
Budafoki \'ut 8., H-1111 Budapest, Hungary}
\author{C.P. Moca}
\affiliation{BME-MTA Exotic Quantum Phases Research Group, Institute of Physics, Budapest University of Technology and Economics, 
Budafoki \'ut 8., H-1111 Budapest, Hungary}
\affiliation{Department of Physics, University of Oradea, 410087, R-Oradea, Romania}
\author{L. Glazman}
\affiliation{Department of Physics, Yale University, New Haven, Connecticut 06520, USA}
\author{G. Zar\'and}
\affiliation{BME-MTA Exotic Quantum Phases Research Group, Institute of Physics, Budapest University of Technology and Economics, 
Budafoki \'ut 8., H-1111 Budapest, Hungary}

\begin{abstract}
We present a detailed theoretical analysis of the Wigner crystal states  in confined semiconducting carbon nanotubes. 
We show by robust scaling arguments as well as by detailed semi-microscopic calculations that the effective exchange interaction  has an SU(4) symmetry, and   can reach values even  as large as $J\sim 100 {\rm \,K}$ in weakly screened, small diameter 
nanotubes, close to  the Wigner crystal - electron liquid crossover. Modeling the nanotube 
carefully
and analyzing the magnetic structure of the inhomogeneous electron crystal, 
we recover the experimentally observed 'phase boundaries'  of Deshpande and Bockrath
[V. V. Deshpande and M. Bockrath, Nature Physics {\bf 4}, 314 (2008)]. 
Spin-orbit coupling only slightly modifies these phase boundaries,  but breaks the spin symmetry down to SU(2)$\times$SU(2),   and  in Wigner molecules it gives rise to interesting excitation spectra, reflecting the underlying 
SU(4) as well as the residual SU(2)$\times$SU(2) symmetries.
\end{abstract}

\maketitle

\section{Introduction}\label{sec:Introduction}

Electrons interacting through simple Coulomb interaction  represent a most fundamental, nevertheless  
challenging interacting quantum system. Apart from dimensionality (${\rm D}$), the behavior of
a Coulomb gas depends  on just two parameters: the temperature $T$, and the  strength of the Coulomb interaction
relative to the electrons' kinetic energy, characterized by the 
dimensionless ratio~\cite{Matveev.2004} 
\beq
r_s = \frac{e^2 m^*}{\epsilon\; \hbar^2 n_e^{1/{\rm D}}}, 
\label{eq:r_s}
\eeq
with $n_e$ denoting the electron density,   $m^*$ the electrons' effective mass, and $\epsilon$ 
the  dielectric constant of the environment through which electrons  interact. 

While at very high temperatures  electrons form a (almost) classical plasma, the behavior of the gas
at low temperatures depends on the specific value of $r_s$. 
At large densities corresponding to  $r_s\ll 1$, the Coulomb interaction plays a minor role in ${\rm D}= 3$ dimensions, and 
 Landau's Fermi liquid state emerges as the temperature is lowered. At small densities ($r_s\gg 1$), 
however, interactions become strong and relevant. In this limit, translational symmetry is broken,  
electrons localize at low temperatures, and  form a Wigner crystal, characterized by 
magnetic ordering~\cite{Wigner.34,BonsallLynnMaradudin.77,CeperleyPRB.2004}. 

While the three-dimensional picture of the previous paragraph  applies also to two 
dimensions~\cite{Peeters.1984, Filinov.2001,Goldman.1990,ChakravartyKivelson.1999,Grimes.1979},  
it fails in one dimension, where quantum fluctuations destroy the long ranged charge order,  remove the phase
transition(s)~\footnote{Several intruding magnetic phases have been proposed, preempting a direct transition between the 
Wigner crystal and the Fermi liquid state.} between the  crystalline and the liquid phases, and replace it by a smooth crossover at some value 
$r_s \approx  r_{1\rm D}^*$~\cite{Schultz,Matveev.2009}.  Though there is no phase transition in one dimension, 
the physical picture is 
 quite different  in the dilute, $r_s\gg r_{1D}^*$,  and dense  regimes,  $r_s\ll r_{1D}^*$. In the Wigner crystal regime, 
 $r_s\gg r_{1D}^*$, the density-density correlation function  reveals charges localized relative to each other, reflected in deep and  long-ranged periodic oscillations~\cite{Piacente.2004, Matveev.2009,Rontani.2010}. 
 In contrast, in the weakly interacting limit, $r_s\ll r_{1D}^*$, these oscillations become weak perturbations on a  non-oscillating 
 background~\footnote{In case of screening, both regimes can be  described as a Luttinger liquid. However, while the charge- and spin velocities  
 are about the same for $r_s\ll r^*$, the spin velocity $c_s$ gets exponentially suppressed compared to the 
 charge velocity  $c_c$ in the Wigner crystal regime,$r_s\gg r^*$.}.
These differences are even more pronounced in a finite system, where charges are typically  pinned 
by some walls or confining potentials, and a  true Wigner crystal structure emerges at small 
densities~\cite{Glazman.1992,Rontani.2010,Guerrero.2014}. 

\begin{figure}[b]
  \includegraphics[width=\columnwidth]{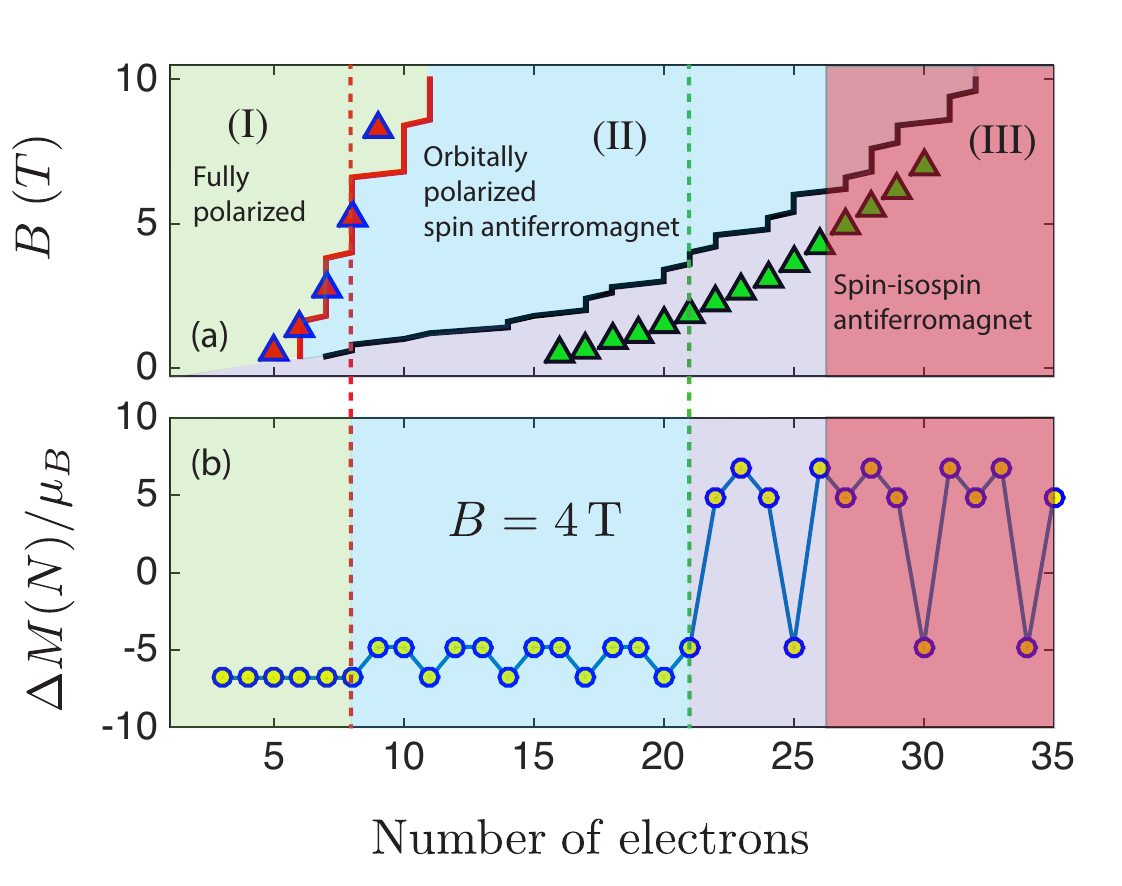}
  \caption{(Color online) (a) 'Phase diagram'
for  an  unscreened nanotube with $\epsilon=2$ and of a radius $R=1.6{\,\rm nm}$,
as a function of particle number $N$, and  external magnetic field $B$. 
Red and black curves denote the phase boundaries based upon our theoretical calculations with spin-orbit 
coupling, $\Delta_{\text{SO}}=0.186\,\rm meV$ ($\Delta_{\text{SO}}\approx 2.1 {\rm K}$). The  fully polarized,  orbitally polarized, and unpolarized states  
are displayed in different colors.  Green and red symbols indicate the boundaries for the experimental 'phase diagram' of Ref.~\cite{Deshpande.2008}. (b)  
Overall magnetization change at charge transitions as a function of $N$ for $B=4\,\rm T$. The vertical dashed lines indicate the  jumps corresponding to the phase boundaries. 
For $N>26$ the Wigner crystal starts to 'melt'. The melted region is indicated by the red shaded areas.}
\label{fig:phase_diagram_soi0}
\end{figure}

In a recent experiment, Deshpande  and Bockrath reported the (indirect) observation of a Wigner crystal state
in a suspended carbon nanotube with a presumably confinement induced gap~\cite{Deshpande.2008, mynote_gap}.  
In a finite magnetic field, they have observed 
 oscillations in the addition energy of holes in a p-type nanotube, and argued that 
only a Wigner crystal picture is able to account for these oscillations systematically 
 (see Fig.~\ref{fig:phase_diagram_soi0} for their 'phase diagram').  More recently, 
 an isolated  two-electron Wigner molecule has been observed in an ultraclean carbon 
 nanotube~\cite{Ilani.2013}. 
 In the latter experiments, the observed level structure has been supported by detailed 'ab initio' 
 calculations, evidencing an exchange splitting much below the single particle level spacing, 
 a clear indication of the Wigner crystal regime. 
 In addition to these experiments, circumstantial   evidence to support the formation of a Wigner crystal 
in one-dimensional wires has also been reported recently by other experimental groups~\cite{Yamamoto.2006, Hew.2009, Yamamoto.2012}.  
It has been suggested  in Ref.~\cite{Yamamoto.2006}, e.g., that  Wigner crystallization  accounts for the negative Coulomb drag 
effect observed  in  coupled  parallel quantum wires.

Although for the experiments  of Ref.~\cite{Deshpande.2008} the Wigner crystal picture seemed to provide a coherent 
explanation,  a sufficiently 'microscopic' theoretical description of the nanotube's Wigner crystal state was missing.  
Furthermore,  in several other experiments~\cite{Meirav.89, Aleshin.2004} a single-particle or 
  mean field scenario, or eventually a  Luttinger liquid picture appears to be sufficient~\cite{Giamarchi, Levitov.2003, Deshpande.2010}. 
The  goal of the present work is   to provide a detailed and quantitative theoretical description of 
an inhomogeneous Wigner crystal in  a  gapped  carbon nanotube, to
investigate its intriguing spin and pseudospin (chirality)  physics, 
and to make a comparison  with the experimental findings of Ref.~\cite{Deshpande.2008}.
For this purpose, we use a bottom-up approach.
 We start out  from  a detailed 
 microscopic modeling of the nanotube, similar to Ref.~\cite{Rontani.2010},  and extract the effective exchange 
 interaction   of two neighboring electrons in the Hartree field of all other electrons from their two-electron spectrum. 
 Exchange processes involve both the spin  ($\sigma$) and chiral spin ($\tau$) of the the crystallized electrons, 
 and an almost perfectly SU(4)-symmetric exchange  interaction is recovered~\cite{Sutherland.1975, Li.1998}.
 We then determine  the positions of the crystallized electrons self-consistently at the classical level 
 and,  having  the separation dependence of the exchange coupling $J$ in hand,  arrive at an 
 effective exchange Hamiltonian, 
 \begin{equation}
\label{eq:su4_hamilton}
	H_{\rm X}=  \frac{1}2  \sum_{i}  J_i\, X^\sigma_{i,i+1} X^\tau_{i,i+1},	
\end{equation}
with the operators $X^\sigma_{i,i+1}$ and $X^\tau_{i,i+1}$ exchanging the spin and the chiral spin of neighboring  electrons
in the crystal and $J_i=J(d_{i,i+1})$.  
Although the exchange Hamiltonian $H_{\rm X}$  possesses  SU(4) symmetry, 
 this symmetry is  broken to  SU(4)$\,\to\,$SU(2)$\times$SU(2) by the spin-orbit (SO) 
 coupling~\cite{Ando.2000, Huertas.2006, Jespersen.2011,Kuemmeth.2008}. 
The  spin-orbit coupling  between the motion of the particles around the tube and their spin 
can be taken into account by the term
\begin{equation}
\label{ham_soi}
	H_{\text{SO}}= -  \frac {1} 2\,\sum_{i=1}^N \,\Delta_{SO} \,\sigma_{i} \tau_{i}\;,
\end{equation}
with $\tau_i$  and $\sigma_i$ denoting the chirality and spin of the $i$th localized electron, respectively, and $\Delta_{\rm SO}$ the spin-orbit splitting.
  We use the effective Hamiltonians Eq.~\eqref{eq:su4_hamilton} and Eq.~\eqref{ham_soi} 
first to construct and classify the low energy  excitations of 
 Wigner molecules, and then to construct the 'phase diagram' of a parabolically confined Wigner crystal
 in a magnetic field by means of an inhomogeneous valence bond approach~\cite{Levitov.2003}.  

 The final 'phase diagram', constructed from the magnetization jumps between different charging states of the nanotube
 is summarized in Fig.~\ref{fig:phase_diagram_soi0} (and discussed in Section~\ref{sec:inhom}). 
 Our theoretical 'phase diagram' compares  astonishingly  well with the experimentally 
determined  phase boundaries in Ref.~\cite{Deshpande.2008}. 
Our calculations rely on just a few parameters, estimated  from the experimental data: 
the radius $R=1.6\, \rm nm$ of the nanotube, yielding  the experimentally reported curvature 
induced gap $E_{g} \sim 220\,\rm meV$~\cite{Kouwenhoven.2015},  the dielectric constant $\epsilon$,  
 the strength of a parabolic  confinement potential $\alpha$,  defined in Eq.~\eqref{eq:conf}   and determined from the addition energy spectra 
(see Ref.~\cite{Kouwenhoven.2006} and Appendix~\ref{app:charging_energy} for details),   and the measured orbital magnetic moment ($g$-factor). 
We thus have one unknown parameter, the dielectric constant $\epsilon$. The value of $\epsilon$ incorporates various screening 
effects including that of  plasmonic excitations and, depending on the specific arrangements and the chirality of the nanotube, can take very different values~\cite{Kozinsky.2006}.  Throughout this paper  we  use 
the value observed in suspended nanotubes~\cite{Miyauchi.2007} and suspended low density graphene~\cite{Guinea.2010},  $\epsilon\approx 2$, incorporating   short distance screening effects. The choice $\epsilon =1 $ would also appear to  be natural~\cite{IlaniPrivate}, however, as we discuss later, 
 this value seems to be inconsistent with the  experimental observations of Ref.~\cite{Deshpande.2008}.

We also find that for these parameters,  supported by the experimental data of Ref.~\cite{Deshpande.2008},
 the   Wigner crystal picture can be appropriate up to around $N\sim 26$ electrons, where the Wigner crystal starts to 'melt'. 
   The  crossover from the Wigner crystal to the electron liquid regime occurs 
 at a crossover value $r_s\approx r_{1D}^*$, which we  estimate based upon  exact diagonalization calculations similar to those in  Ref.~\cite{Rontani.2010} to be $r_{1D}^*\approx 3.3$ (see also Section~\ref{sec:scaling}).  
At this crossover value of  $r_s\approx r_{1D}^*$ (the boundary of the Wigner crystal regime) the 
 one-dimensional density $n_e^*$ of the electrons (holes) and their exchange 
 coupling $J^*$   scale as 
 \beq
 J^*\sim {m^*}/{\epsilon^2}\;, \phantom{nnn}  n_e^*\sim {m^*}/{\epsilon}\;.
 \label{eq:scaling}
 \eeq 
The effective mass of a semiconducting nanotube
depends sensitively  on the radius $R$ of the nanotube, 
 $m^*\sim 1/R$. Therefore, the precise density range of Wigner crystal behavior as well as the energy scale  of the 
exchange interaction in the Wigner crystal regime are very sensitive to the 
\emph{radius} $R$ of the tube as well as the precise \emph{value of $\epsilon$}.  For  a nanotube of radius $R = 3\,{\rm nm}$
and $\epsilon = 4.5$, detailed estimates  yielded a relatively small exchange coupling
$J^*_{\rm }\approx 10 \,{\rm K}$ and  a large crossover separation $d^*_{\rm } = 1/n_e^*=50\,{\rm nm}$~\cite{Ilani.2013}.  
On the other hand,  for a nanotube of radius $R=1.6 \,{\rm nm} $ 
(yielding a band gap close to the one reported  in Ref.~\cite{Deshpande.2008}),
and of moderate screening, $\epsilon\approx 2$, 
Eq.~\eqref{eq:scaling} immediately yields a surprisingly large 'crossover' exchange coupling along with 
 a small  carrier separation,
\beq
J^*\sim 95 \,{\rm K}, \phantom{nnnn}  d^* \sim  11.8 \, {\rm nm}.  
\eeq
We thus conclude that  the first  transition line in Fig.~\ref{fig:phase_diagram_soi0} occurs well within the Wigner crystal regime, supporting  the interpretation of the authors of Ref.~\cite{Deshpande.2008}, while the second transition line reaches into the melted region for high magnetic fields. 
We  emphasize, however,  that stronger screening by the environment, $\epsilon\gtrsim 4$ 
quickly reduces  $J^*$ to the few Kelvin range, and increases simultaneously the characteristic 
separation of particles to $ d^* \sim  80-100 \, {\rm nm}$.

The rest of the paper is structured as follows: 
In Sec.~\ref{sec:EffectiveHamiltonian} we determine the effective exchange interaction between two neighboring electrons in the Wigner crystal state and  construct the effective spin Hamiltonian of a  
one dimensional electron crystal in the nanotube. In Sec.~\ref{sec:WignerMolecules}
we analyze the magnetic excitations of small Wigner molecules and show how spin-orbit interaction 
breaks the SU(4) symmetrical spectrum to SU(2)$\times$SU(2) multiplets. 
In Sec.~\ref{sec:inhom} we investigate the spin structure of confined inhomogeneous 
nanotubes in an external magnetic field using a fermionic valence-bond calculation.
Finally, before concluding, in Sec.~\ref{sec:scaling} 
we discuss the limitations and consistency of the Wigner crystal approach and 
present the simple scaling arguments, leading to the relations Eq.~\eqref{eq:scaling}. 
Four appendices explain  some useful details of these calculations.

\section{Construction  of the effective Hamiltonian}
\label{sec:EffectiveHamiltonian}

\subsection{Derivation of the exchange coupling }
\label{sub:DerivationExchange}

To determine the exchange coupling in the Wigner crystal regime, we use a bottom-up approach. First, we 
model the interaction of two neighboring electrons in detail by semi-microscopic calculations~\cite{Rontani.2010}, 
and extract their exchange interaction from the two-particle excitation spectrum.
We find that the exchange interaction is quite accurately given by a semiclassical expression, 
similar to the ones used in Refs.~\cite{Hausler.1995,MatveevPRB.2004}.

In  small diameter semiconducting nanotubes discussed here, 
and at scales larger than the  atomic scale,  the motion of the interacting $N$ electrons (holes) 
 is well described  in terms of the lowest conduction (or highest valence) bands and the   corresponding 
  effective Hamiltonian 
\begin{equation}
 \label{eq:nelectron}
 H=-\frac{\hbar^2}{2m^*}\sum_{i=1}^N\,\frac{\partial^2}{\partial z_i^2}+\sum_{i<j}U(z_{ij},\varphi_{ij}), 
\end{equation}
with $z_i$ and $\varphi_i$ denoting the particles' cylindrical  coordinates (see also Appendix~\ref{app:ECP})~\footnote{Notice that after  projection to the lowest conduction band, the kinetic energy  contains only the coordinates $z_i$, but the wave functions still have an angular dependence, as dictated by the  chirality quantum number. The $\varphi$-dependence therefore still appears in the interaction-part of  Eq.~\eqref{eq:nelectron}.}.
The effective mass $m^*$ here is simply related to the gap of the nanotube as $E_g=2m^*c^2$ with 
$c \approx 8 \times 10^{5} {\rm m/s}$ the Fermi velocity of graphene. 
For the sake of simplicity and concreteness, here we  discuss  electron-doped small radius semiconducting nanotubes, where the  gap is mostly  due to radial confinement and $E_g\approx 2\hbar c /3 R$~\cite{Kouwenhoven.2015}, but our discussion 
carries over  with trivial modifications to hole-doping and nanotubes with strain or curvature induced gaps, too.  
In these latter cases, however, particles are typically lighter and it is harder to reach the Wigner crystal regime 
experimentally.

The Coulomb interaction 
\begin{equation}
\label{eq:u}
U(z,\varphi)=\frac{e^2}{\epsilon}\,\frac{1}{\sqrt{z^2+2R^2(1-\cos\,\varphi)}}\, ,
\end{equation}
 depends just on the  
distance  between particles and thus $z\to z_i-z_j$ and $\varphi\to  \varphi_i-\varphi_j$
~\footnote{A microscopic cut-off is usually also introduced to regularize it
 on the atomic scale (see Appendix~\ref{app:ECP}).}. The  dielectric constant  $\epsilon$ 
  depends crucially on the 
way the nanotube is prepared and contacted; for a nanotube laid over a typical semiconductor 
$\epsilon \sim 5-7$ seems to be a reasonable estimate~\cite{Homma.2009}, while in suspended nanotubes it may get close to the vacuum 
value, and $\epsilon\approx 2-3$ or possibly even smaller values seems to be a realistic choice~\cite{IlaniPrivate,Homma.2009}. 


\begin{figure}[t]
  \includegraphics[width=.45\textwidth]{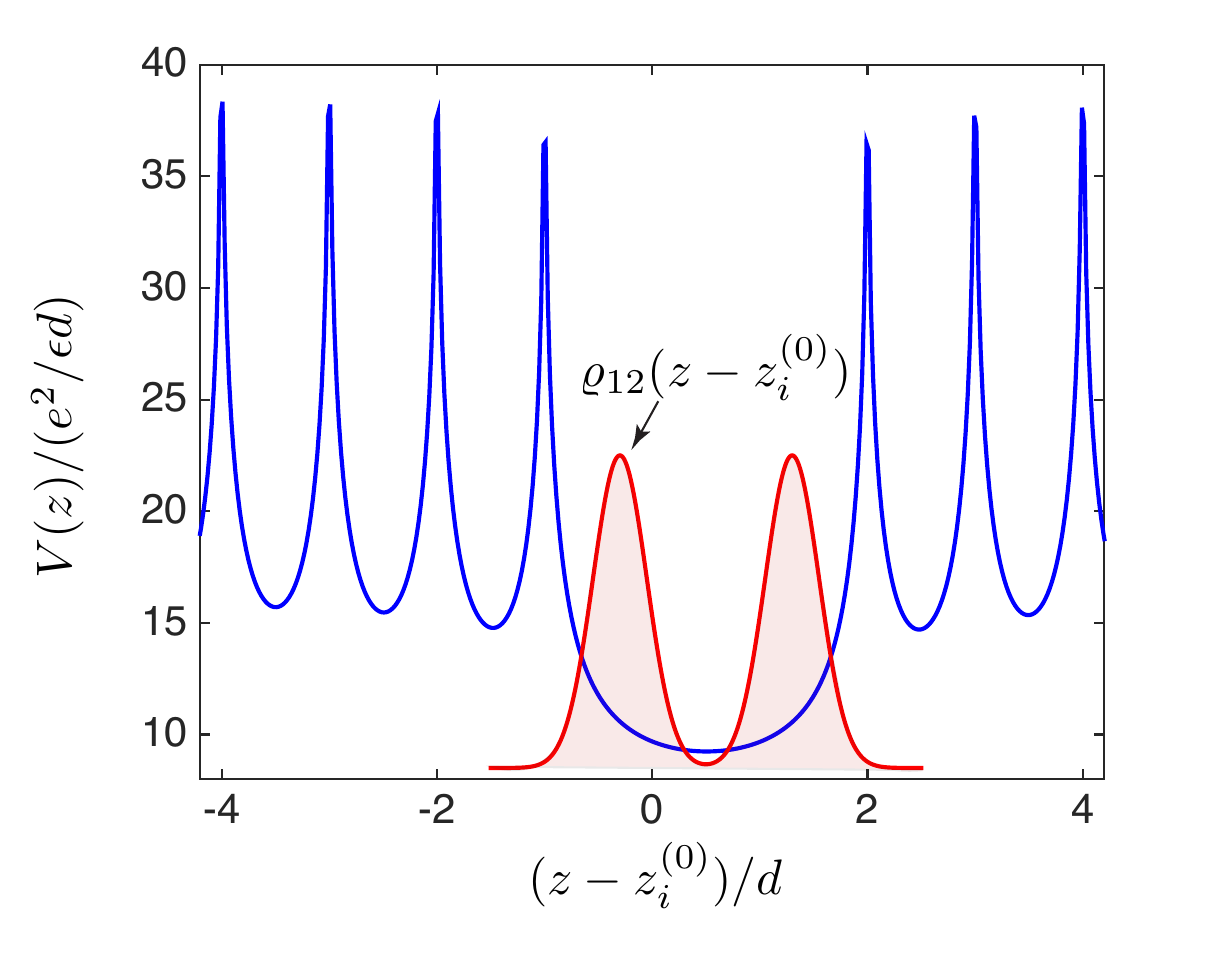}
  \caption{(Color online) Sketch of the effective potential, $V(z)$ (blue), and the single particle density (red), as obtained from the solution 
 of the two body problem, Eq.~\eqref{eq:ham-2p}. d is the distance between two neighboring electrons.
}
\label{fig:vij}
\end{figure}

Consider now two neighboring particles 
in the Wigner crystal regime, 
moving in the  Hartree field of the other particles, and interacting with each other as described by the 
effective Hamiltonian, 
\begin{eqnarray}
   H^{(2)}&=&-\frac{\hbar^2}{2m^*}\,\left( \frac{\partial^2}{\partial z_1^2} + \frac{\partial^2}{\partial z_2^2}\right)+V(z_1)+V(z_2)
   \nonumber
   \\
   &&+ U(z_{12},\varphi_{12}).  
   \label{eq:ham-2p}
\end{eqnarray}
The  Hartree potential $V(z)$, displayed in 
Fig.~\ref{fig:vij},  is well approximated in the Wigner crystal regime  as  
\begin{equation*}
V(z)\approx \sum_{j\neq 1, 2}  U_0(z-z_j^{(0)}),
\end{equation*}
with the $z_j^{(0)}$ denoting the classically obtained locations of the other particles, and 
$U_0(z)$ the angular averaged Coulomb interaction, 
\begin{equation}
\label{eq:h12-offdiag}
    \begin{split}
U_0(z) \equiv \int_0^{2\pi}\frac{\dd\varphi }{2\pi}\,U(z,\varphi).
  \end{split}
\end{equation}

The angular dependence of the wave function is determined by the isospin (chirality) $\tau=\pm$ of the electrons, which  also enters the  two-particle wave function  as
$\psi = \Phi^{\sigma_1,\sigma_2}_{\tau_1,\tau_2}(z_1,z_2)\,e^{i Q (\tau_1\varphi_1 + \tau_2\varphi_2)}$, with the angular momentum 
$Q$ determined by the chirality of the tube~\footnote{For a gapped zig-zag nanotube of chirality $(3k+\nu,0)$, 
considered here,
$Q= 2k + \nu$ ($\nu=\pm1$).} (details of the derivation are provided in Appendix~\ref{sec:hilbert-space}).
The interaction term in Eq.~\eqref{eq:ham-2p}  preserves the total isospin of the two interacting particles: $\tau_1+\tau_2=\tau_1'+\tau_2'$, but the matrix elements of $H^{(2)}$ 
do depend on the relative values of $\tau_1$ and $\tau_1'$. Nevertheless, while for $\tau_1=\tau_1'$ integration over the angles yields
the effective  interaction, $U_{\tau_1\tau_2}^{\tau_1\tau_2}(z_{12}) = U_0(z_{12})$, the off-diagonal matrix elements of the potential, 
$U^{\tau_2 \tau_1}_{\tau_1\tau_2} $ are found to be  several orders of magnitude smaller than $U_0$ for $\tau_1\ne\tau_2$
due to  rapid oscillations of the wave functions~\cite{Rontani.2010}.
Therefore, to a very good accuracy, the two-body Hamiltonian is diagonal in the spin and isospin quantum 
numbers,  and the corresponding Schr\"odinger equation reads 
\beq
E\,\Phi^{\sigma_1,\sigma_2}_{\tau_1,\tau_2} \approx  \left[-\frac{\hbar^2}{2m^*}\left( \frac{\partial^2}{\partial z_1^2}  + \frac{\partial^2}{\partial z_2^2}\right)
+\,U_{\rm tot}(z_1,z_2) \right ]\Phi^{\sigma_1,\sigma_2}_{\tau_1,\tau_2}
\label{eq:twobody}
\eeq 
with  $U_{\rm tot}= V(z_1)+V(z_2)+U_0(z_{12})$ the total two-body potential. 

\begin{figure}
  \includegraphics[width=.4\textwidth]{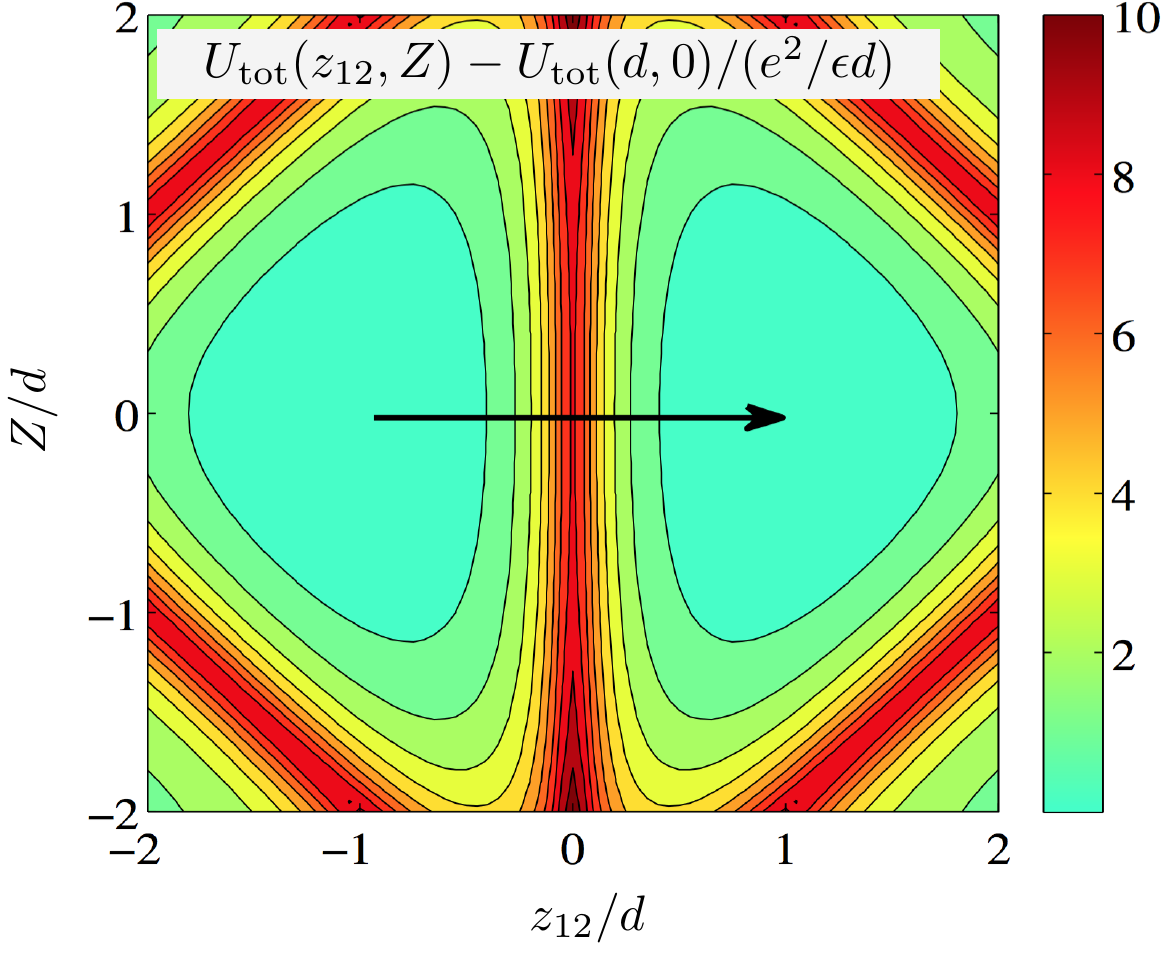}
   \caption{(Color online) Contour plot of the background potential  $U_\text{tot}$ in Eq.~\eqref{eq:twobody} in terms of relative and center of mass 
   coordinates. Tunneling between the two minima ($z_{12}=\pm d,Z=0$) gives rise to the exchange splitting $J$. 
   The black arrow indicates the semiclassical tunneling path used here.
   }
\label{fig:Utot}
\end{figure}

The  chiral and spin quantum numbers are tied together with the 
orbital part of the wave function by Pauli's principle. Spatially-even solutions of \eqref{eq:twobody} imply 
antisymmetry under exchanges $(\sigma_1,\tau_1)\leftrightarrow (\sigma_2,\tau_2)$, while  
 odd solutions must be  symmetric under them. 
Therefore,  the spectrum of the lowest 16 eigenstates 
of  \eqref{eq:twobody} is reproduced  by the effective spin exchange Hamiltonian 
\begin{align}
\label{eq:exchange-op}
 H_J=\frac{J}{2}\cdot \xs\xt,  
\end{align}
with $J$ denoting the splitting of the 6-fold degenerate  ground state and the 10-fold degenerate first excited multiplet. 
These large degeneracies are due to the (approximate)  SU(4) symmetry of the exchange 
interaction.

The splitting $J$ can be extracted by diagonalizing the two-body Hamiltonian~\cite{MultiparticleNote} or, 
alternatively, in  the Wigner crystal regime one can determine it with a remarkable 
$\sim 15 \%$ accuracy by means of a semiclassical approach (see Appendix~\ref{app:WM}). 
Displaying the two-body potential $U_{\text{tot}}$  in terms of the 
relative and  center of mass coordinates $z_{12}$ and $Z=(z_1+z_2)/2$,  
we notice that  the two particles move in a double-well potential
(see Fig. ~\ref{fig:Utot}).  Tunneling processes along the tunneling path 
indicated in Fig.~\ref{fig:Utot} lift the degeneracy of left and right states associated with the minima of 
$U_{\text{tot}}$, and give rise to the exchange splitting  $J$.  

\begin{figure}[t]
	\includegraphics[width=.5\textwidth]{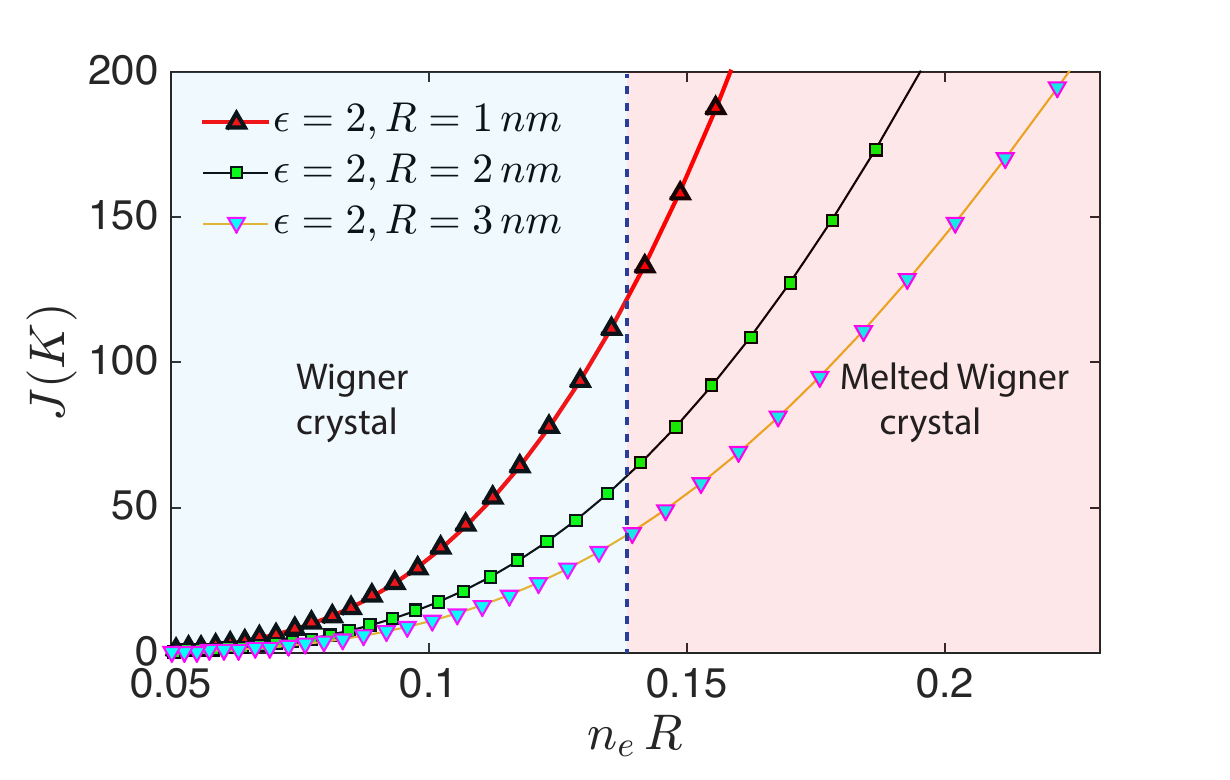}
	\caption{(Color online) The effective SU(4) spin-isospin exchange coupling $J$ as a function of $n_e\cdot R$ for various nanotube radii
	with $\epsilon=2$. The vertical dashed line shows the limit of the Wigner crystal regime.}
\label{fig:delta}
\end{figure}

The couplings $J$ are displayed  in Fig.~\ref{fig:delta} as a function of $n_e\cdot R$
for several experimentally relevant nanotube radii and $\epsilon=2$. 
 In these density units, the boundary of the 
 Wigner crystal is at $(n_e\cdot R)^*\approx 0.135$
 independently of  the radius of the nanotube (see Section~\ref{sec:scaling}).
For small $R\approx 1 \,{\rm nm}$ radius nanotubes we find that the exchange coupling can be as high as 
$J\sim 100 \,{\rm K}$ before the crystal starts to 'melt'.  This value can be even higher in case $\epsilon $ is closer to 1.


\subsection{Effective spin Hamiltonian of a carbon nanotube Wigner crystal}

Having determined the exchange coupling $J(n_e)$ in  a homogeneous gas of electrons, we are 
now in a position to construct an effective Wigner crystal Hamiltonian. Assuming that the density of the electron or hole gas 
 changes relatively slowly on the scale of typical particle-particle separations and that 
 charges are  pinned by some confining potential,   we can neglect charge fluctuations and 
approximate the exchange coupling of two neighboring particles as $J(d)$, with $d=n_e^{-1}$ denoting their separation.  
We thereby arrive at the effective exchange Hamiltonian $H_X$ in Eq.~\eqref{eq:su4_hamilton}. 

The spin-orbit coupling, Eq.~\eqref{ham_soi},  does not influence the exchange coupling substantially, and we have therefore neglected it in the previous subsection. However,  deeper in the Wigner crystal regime (or in 
screened nanotubes) it can be comparable with  $J$ or larger, and can influence the spin state of the nanotube essentially~\cite{Moca_unpublished}. 
The value of the coupling $\Delta_{\rm SO}$ is roughly inversely proportional 
to the radius of the tube, $\Delta_{\rm SO}\cdot R \sim 0.3~\text{meV}\cdot\text{nm}$, though sample to sample fluctuations can be 
 large~\cite{Kouwenhoven.2015}. 
As already mentioned in the Introduction, $H_{\text{SO}}$ breaks the SU(4) symmetry of the exchange Hamiltonian
Eq.~\eqref{eq:su4_hamilton} down to $\text{SU(2)}\times \text{SU(2)}$. 
However, $\Delta_{\rm SO}$ turns out to be relatively small in the crossover regime compared to 
the exchange coupling in poorly screened ($\epsilon\le 2$) nanotubes.  
For a nanotube of radius $R=1.6 {\rm \, nm}$ and $\epsilon =2$, e.g., 
$\Delta_{\rm SO}\approx 2.1 \,{\rm K}$~\cite{Kouwenhoven.2015}, which 
is about a factor $\sim 40$ smaller than the  exchange coupling  in the 
crossover regime. 

The symmetry of the Wigner crystal state is further reduced in the presence of an external magnetic field.
Here we  focus on the simple case of a magnetic field   parallel to the axis of the nanotube, when
\begin{equation}
\label{eq:zeeman}
	H_{B}=\sum_{i}\,\mu_{B} B\left(\frac 1 2 \; g_s \sigma_{i} + g_{\text{orb}}\tau_{i}\right).
\end{equation}
with $g_s\approx2$ and $\mu_B = 0.057 {\rm \,meV/T}$ the Bohr magneton. 
Thereby the ${\rm SU(2)}\times {\rm SU(2) }$ symmetry of the crystal
is further reduced to $ \text{U(1)}\times \text{U(1)}$.
 
For an infinite nanotube the orbital $g$-factor $g_{\text{orb}}$ can be estimated as 
$g_{\text{orb}}   \approx 7\cdot R \,{\rm [nm]}$ , however, this value  can be substantially reduced  by 
confinement~\cite{Jespersen.2011}.  Therefore, for the nanotube  in Fig.~\ref{fig:phase_diagram_soi0}
we have used the experimentally extracted $g$-factor, 
$g_{\text{orb}}\approx 5.8$, {corresponding to $\mu_{\rm orb}= g_{\rm orb}\,\mu_{B}\approx 0.33\, \rm {meV/T}$}~\cite{Deshpande.2008},
yielding indeed good agreement for the 'phase boundaries' in Fig.~\ref{fig:phase_diagram_soi0}.

\section{Wigner molecules}
\label{sec:WignerMolecules}

\begin{figure}[tbhp!]
\includegraphics[width=0.5\textwidth]{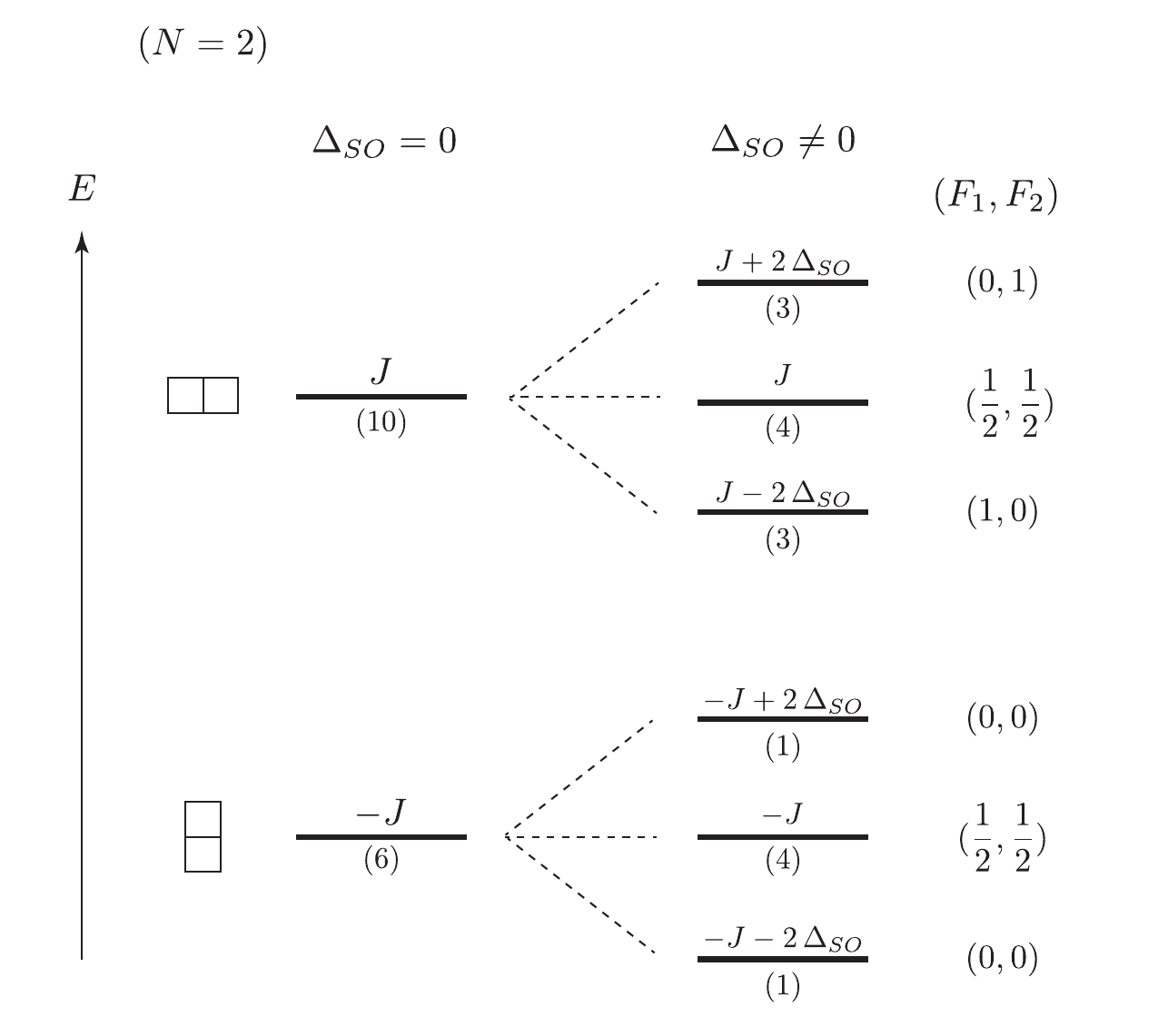}
\caption{
Energy levels of an $N=2$ Wigner molecule, as obtained from exact diagonalization of the effective Hamiltonian. 
On the l.h.s., the $\Delta_{\rm SO}=0$ spectrum is shown, with states classified by SU(4)  representations, indicated by the Young-tableaux.
For $\Delta_{\rm SO}\ne 0$  states are classified in terms of  the residual SU(2)$\times$SU(2)
symmetry.}  
\label{fig:ed}
\end{figure}

For small systems of $N=2$ and 3 electrons (holes)  Wigner molecules form, and we can diagonalize  
the effective Hamiltonians Eqs.~\eqref{eq:su4_hamilton} and \eqref{ham_soi}  analytically to obtain  their spin  excitation spectrum.
 For $\Delta_{\rm SO}=0$, the spectrum is organized into SU(4) multiplets characterized by Young tableaux.  For a $2$ particle molecule, e.g., 
the lowest $4^2=16$ states  are organized into a 6-fold degenerate antisymmetric ground state multiplet 
 and a 10-fold degenerate symmetrical excited state (see Fig.~\ref{fig:ed}). 

These highly degenerate multiplets are split  for $\Delta_{\rm SO}\ne0$, and can be 
 classified by the residual SU(2)$\times$SU(2) symmetries, with  their generators 
 $F_1$ and $F_2$  inducing internal rotations within the $\tau_i\,, \sigma_i = \pm 1$ subspaces. 
In terms of these latter, the 6-fold degenerate  ground state is split to  two $(F_1, F_2) = (0,0)$ singlet states,  
 and a fourfold degenerate   $(F_1, F_2) = (1/2,1/2)$ excited multiplet. 
 In contrast, the 10-times degenerate excited state splits into  
a 4-fold degenerate  $(F_1, F_2) = (1/2,1/2)$ multiplet, and two three-fold degenerate multiplets,
  $(F_1, F_2) = (0,1)$ and a $(F_1, F_2) = (1,0)$ for $\Delta_{\rm SO}\ne 0$.

\begin{figure}[tbhp!]
\includegraphics[width=0.5\textwidth]{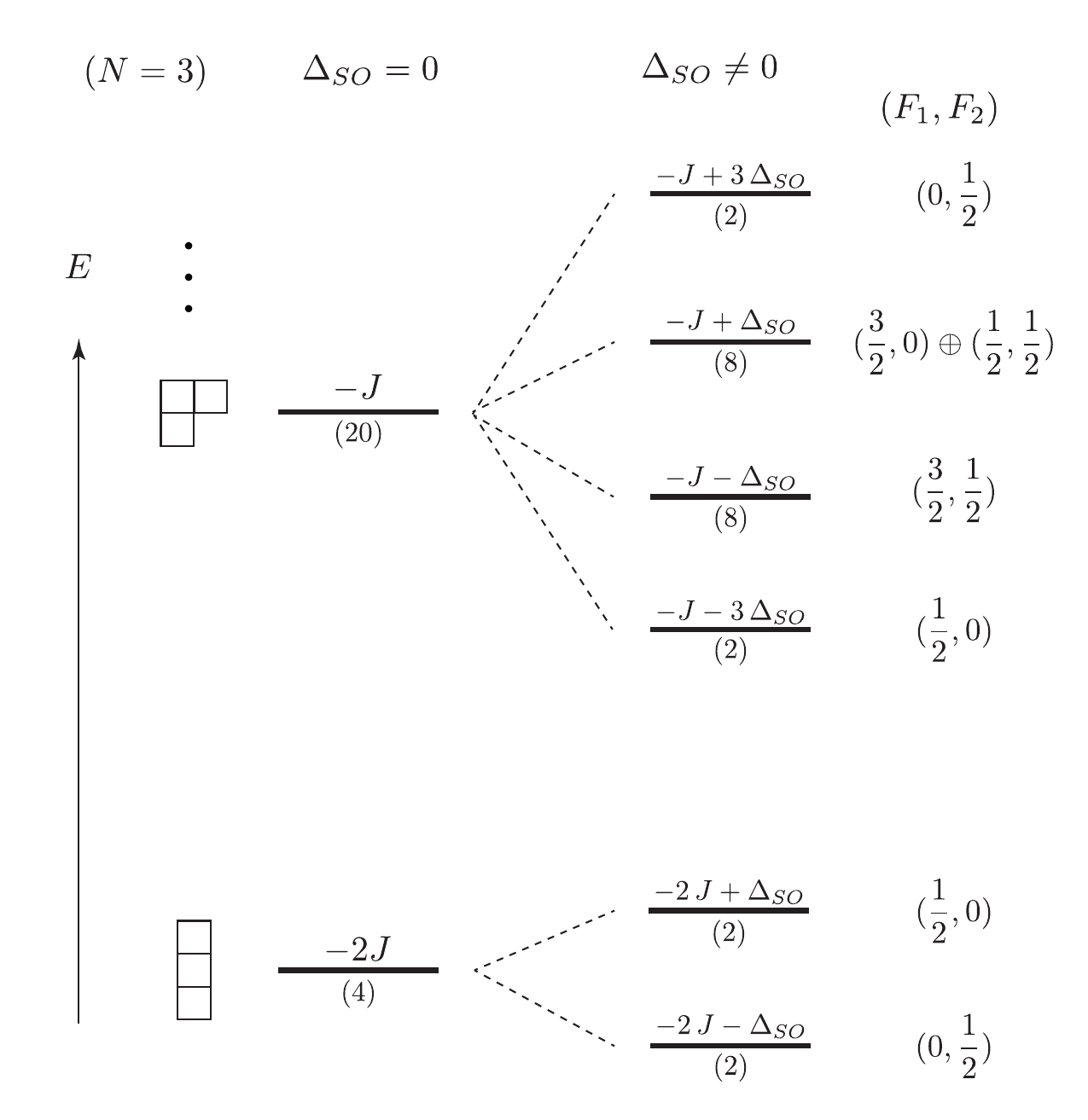}
\caption{Energy levels of a $N=3$ Wigner molecule. 
Only the first two SU(4) multiplets and their spin-orbit coupling induced splittings are  shown. }
\label{fig:ed3}
\end{figure}

 Injecting a third  carrier into the nanotube, an $N=3$-particle Wigner molecule forms.
 The  Hilbert space of low-lying spin excitations is then $64$-dimensional, these 64 states are, however, organized into 
 just four  SU(4) multiplets in the absence of $\Delta_{\rm SO}$: 
the 4-fold degenerate ground state is  again completely antisymmetric in the united spin-isospin space, 
while  excited multiplets have mixed symmetries and 
are all 20-fold degenerate. Similar to 
the case of the $N=2$ molecule, these states can be classified in terms of $(F_1, F_2) $, too,
 and their spin-orbit coupling induced splitting  and their energy can be exactly determined 
 with group theoretical methods (see Fig.~\ref{fig:ed3}).

Injecting yet another carrier, an $N=4$ Wigner molecule forms. In this case, 
even if we assume  that the Wigner molecule  is symmetrical relative  to its center, 
  two distinct couplings  need to be introduced, one for the central bond ($J$), and another  
one for the two side bonds ($J'$).  The excitation spectrum cannot be determined analytically 
in this case, but the ground state is found to be  an SU(4) singlet, as expected.
The approximately SU(4) symmetrical molecular states 
should be visible in the molecules' co-tunneling spectrum,
and 
may lead to interesting quantum states when coupled to 
electrodes (see Summary and Conclusions)~\cite{ZarandGoldhaber2003,Kouwenhoven2005}. The SU(4) spin, coupled 
to two (SU(4)) Luttinger liquids, e.g., could give rise to an SU(4) two-channel 
Kondo state, characterized by an anomalous scaling dimension $\Delta = 2/3$~\cite{LUDWIG1994}. Coupling the SU(4) Wigner 
molecule to side electrodes would, on the other hand, lead to an SU(4) Fermi liquid state~\footnote{The exponent $\Delta$ refers to the case of the SU(4)$\times$SU(2) model with Fermi liquid leads. 
This exponent is, however, expected to remain unchanged  in the Luttinger liquid  case, too, since correlators 
(i.e., spin excitations)  in the SU(4) spin sector of the leads are  expected to remain 
unaffected by the Luttinger parameter $K$ of the charge sector.}.
The higher dimensional SU(4) spins may give rise to exotic underscreened Kondo states. 

\section{Wigner crystal state in a parabolically confined nanotube}
\label{sec:inhom}

We now turn to the experimental setup  of Ref.~\cite{Deshpande.2008}, where the nanotube was attached 
to source, drain and gate electrodes. In this case, 
the attached gate electrodes can produce Schottky barriers at the ends of the nanotube, and  charges accumulated 
there are expected to create  a smooth external, approximately parabolic 
confinement  potential for the charge carriers:
\begin{equation}
\label{eq:conf}
	V_{\text{conf}}\approx \frac12 \alpha \,z^{2}. 
\end{equation}

The depth of the potential, $\alpha = m^*\omega_0^2$ can  be estimated from 
the measured charging energy  of the nanotube (see Appendix~\ref{app:charging_energy} for details).
Throughout the present work we consider  $\alpha\approx 0.015\,\rm meV/nm^{2}$,
which for $\epsilon= 2$ corresponds to a charging energy of $U\approx 15\,{\rm meV}$, in rough agreement with the 
value reported  in Ref.~\cite{Deshpande.2008}, $U\approx 10-12\,{\rm meV}$.
For a $R=1.6{\rm \,nm}$ nanotube this corresponds to  a confinement energy $\hbar\,\omega_0 \approx 6.25\, \rm meV$. 

\begin{figure}
\includegraphics[width=.45\textwidth]{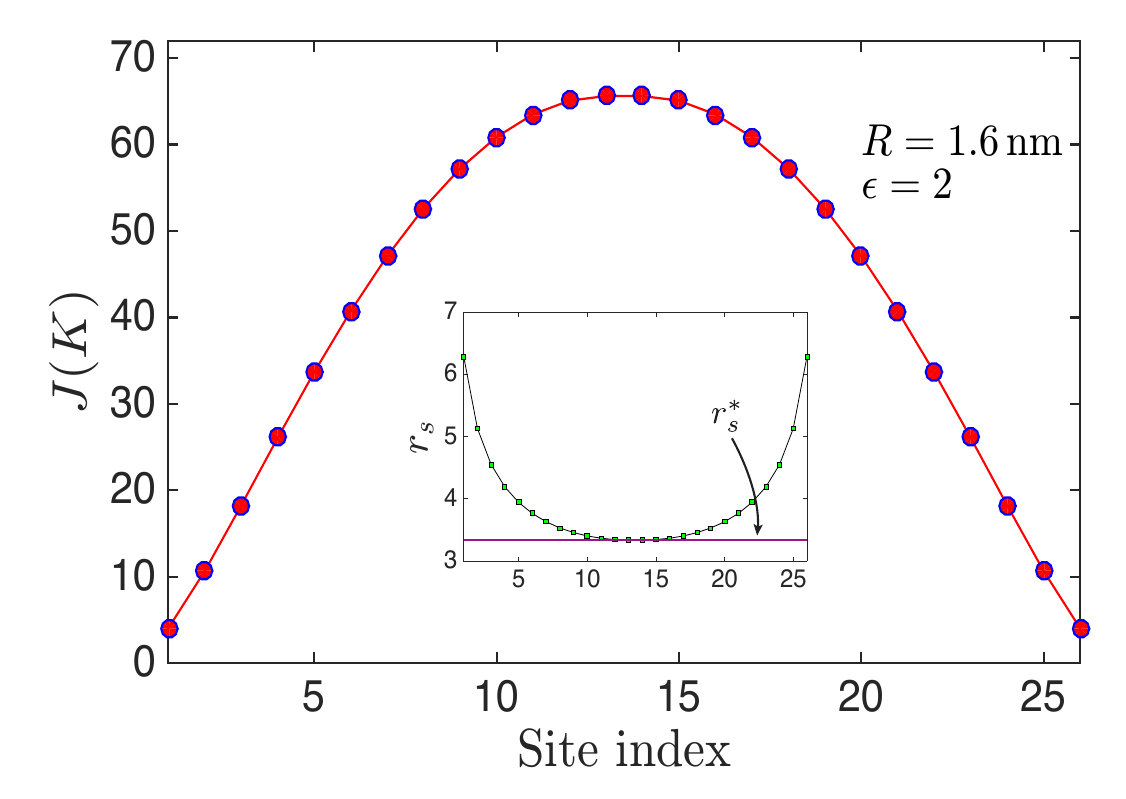}
\caption{(Color online) Site dependent exchange couplings for a nanotube with $\epsilon = 2$ and $R=1.6\,\rm nm$ when $N=26$ electrons are confined inside. The inset shows the dimensionless parameter
  $r_{s}$ as a function of the site index. Notice that  the stability 
   condition $r_{s}\gtrapprox r_{s}^{*}\simeq 3.3$ is fulfilled, and the nanotube is in the Wigner crystal regime, though for larger values 
   of $N$ the Wigner crystal is expected to melt at the center of the nanotube. 
}
\label{fig:dij,jij}
\end{figure}

In a parabolic confinement, the gas of particles is denser at the central region of the tube,
and correspondingly, the exchange coupling $J_i$ is larger there, and decreases towards the ends of the tube. 
 In the experiments the  gate voltage is varied, and  the number of the charge carrier increases one by one 
 until the Wigner crystal gradually melts at the center of the tube. 
 For each $N$ we therefore need to determine   the dynamical rearrangement of the particles and,
  accordingly, a new distribution of the  couplings $J_i$.  To this end, we minimized the classical Coulomb energy of the particles in the confining potential, 
$$
E =   \sum_{i=1}^N \frac {m^*\omega_0^2} 2 z_i^2 + \sum_{i<j}^N\frac {e^2}{\epsilon |z_i-z_j| }\;.
$$  
  
  Fig.~\ref{fig:dij,jij} shows the spatial dependence of 
 the inhomogeneous couplings and the position dependent interaction parameter $r_s$ for  $N=26$ 
 charged particles in a nanotube 
   of $\epsilon=2$ and $R=1.6~\rm nm$.
Remarkably, $r_s$ remains above the crossover value $r_{s}^{*} \approx 3.3$ even at the center of the tube, where the exchange coupling is  as large as $J\sim 65 \, {\rm K}$, and the whole nanotube is in the Wigner crystal regime. 
For larger values,  $N>26$, however, $r_s$ becomes smaller that  $r_s^*$ at the center of the nanotube, and 
the core of the Wigner crystal   melts. 

To determine   the spin state of the relaxed Wigner crystal, we performed 
 a valence bond mean field calculation. We  first grouped
  the spin and isospin variables  as $(\sigma,\tau)\to \alabel$,
 and  rewrote the exchange interaction in a readily SU(4) invariant form  as 
\begin{equation}
\label{hom_hamilton}
	H_X=\sum_{i}\sum_{\alabel\blabel}\frac {J_{i}}2\, c^{\dagger}_{i\alabel}c_{i\blabel}c^{\dagger}_{i+1\blabel}c_{i+1\alabel}\equiv{}
		\sum_{i}\sum_{\alabel\blabel} \frac {J_{i}}2
		\, S_{i}^{\alabel\blabel}S_{i+1}^{\blabel\alabel},
\end{equation}
with the fermionic operators $c^{\dagger}_{i\alabel}$ ($c_{i\alabel}$) creating (annihilating) a carrier  on site $i$, 
and satisfying the constraint $\sum_{\alabel}c^{\dagger}_{i\alabel}c_{i\alabel}=1$ at each site.
The  operators $S_i^{\alabel\blabel}=c^{\dagger}_{i\alabel}c_{i\blabel}$
obey the  SU(N) commutation relations $\left[S_i^{\alabel\blabel},S_j^{\alabel'\blabel'} \right]=\delta_{i,j} (\delta_{\alabel\alabel'} S_i^{\blabel\blabel'} - \delta_{\alabel\blabel'}S_i^{\alabel'\blabel})$. The terms 
$H_{\rm SO}$ and $H_B$ can also be expressed in terms of the SU(4) spin operators, $S_i^{\alabel\blabel}$, and 
obviously break the SU(4) symmetry of $H_X$. Notice that the Hamiltonian 
\eqref{hom_hamilton} is also invariant under local gauge transformations, 
$c_{j\alabel}\to c_{j\alabel} e^{i\theta_j}$, reflecting local particle number conservation.

In the presence of the terms $H_{\rm SO}$ and $H_B$, the inhomogeneity in the exchange coupling plays a crucial 
role.  For longer Wigner chains,  the exchange coupling $J_i$ is very strong  at  the center of the chain, while at the wings it can be smaller by two or more orders of magnitude.
Therefore, in the presence of an external magnetic field, 
the nanotube may be phase separated, with  wings  polarized in isospin 
and spin space, and  the center of the chain  remaining in  an approximately SU(4) antiferromagnetic 
N\'eel state. For smaller external magnetic fields all three phases can be present in the nanotube: 
a spin-isospin polarized state at the wings, where the field-induced Zeeman and orbital splitting are much larger than the exchange coupling, an antiferrmagnetic state at the center, where the exchange coupling dominates over the magnetic field, and  an orbitally polarized spin antiferromagnet between them. 
  
  This competition  of $H_{\rm SO}$ and $H_B$ leads to the nontrivial 'phase diagram' shown 
  in Fig.~\ref{fig:phase_diagram_soi0}, constructed in  Ref.~\cite{Deshpande.2008}
by investigating  the change in  total magnetization while adding an extra particle in a magnetic field, 
$\Delta M(N) = M(N+1)- M(N)$, a quantity which can be directly extracted through transport measurements 
from the shift of the  Coulomb blockade peaks in a magnetic field.
 To understand this 'phase diagram' we must keep in mind that 
the Wigner crystal is inhomogeneous and  therefore the density and the exchange coupling are both the largest 
at the center of the nanotube. Consider now adding particles to the nanotube in a fixed external magnetic field. 
For small particle numbers, the exchange coupling remains small even at the center of the nanotube, 
and both the spins and the  isospins of the entering particles are found to be completely polarized (phase I).
 For increasing particle numbers, however, the density and thus the exchange coupling 
at the center start to become larger, and the latter exceeds the spin splitting of the SU(4) spins, induced by the external magnetic field, but remains still smaller than the orbital splitting, due to the difference in the electronic and orbital 
$g$-factors. As a consequence, new particles  enter  with alternating spins but polarized isospin (isospin polarized spin antiferromagnet, phase II). Finally, for even larger particle numbers, the exchange coupling at the center of the tube  
dominates, and electrons enter with alternating spin and orbital spins (spin-isospin antiferromagnet, phase III). 
Notice that the three regions in Fig.~\ref{fig:phase_diagram_soi0} are not true 'phases'. In region II, e.g., the nanotube 
hosts two magnetic phases:  an orbitally polarized 
spin antiferromagnet at the center, and regions of fully polarized spins and isospins on the wings.

To analyze  the spin-isospin configurations of  the confined Wigner crystal theoretically, 
we made use of  a self-consistent valence bond approach, whereby we decouple  
the exchange term  assuming a finite $Q_i=\langle \sum_{\alabel}c^{\dagger}_{i+1\alabel}c_{i\alabel}\rangle$
to obtain 
\begin{align}
	\label{Ham_MF}
		{H}_\mathrm{MF}=&-\sum_{i\alabel} J_{i}\,\left(Q_{i}\cdot c^{\dagger}_{i+1\alabel}c_{i\alabel}+\hc\right) \nonumber  \\ & +\sum_{i\alabel}\,\mu_{i}\cdot c^{\dagger}_{i\alabel}c_{i\alabel}
		+H_{\text{SO}}+H_{\text{B}} .
\end{align}
Here the Lagrange multipliers $\mu_i$ ensure that particle number is conserved on the average 
at each site and, similar to the $Q_i$, must be determined self-consistently.
Notice that, by the local gauge invariance of \eqref{hom_hamilton},  the $Q_i$ are not uniquely defined, and 
the energies of the ground states of $H_\text{MF}$ remain invariant  under the transformation 
$Q_i \rightarrow Q_i \e^{\text{i}(\theta_{i}-\theta_{i+1} ) }$. This simple mean field approach captures surprisingly well 
the properties of SU(4) and SU(2) antiferromagnets and, according to our findings,  can also account for the 
'phase diagram' of the carbon nanotube Wigner crystal.  


The 'phase diagram'  represented in Fig.~\ref{fig:phase_diagram_soi0} has been determined 
by performing self-consistent calculations for each magnetic field and for up to $N=35$ particles. 
 As already stated in the Introduction, apart from $\epsilon$, which we 
have set to $\epsilon=2$ to agree with the values reported so far in suspended graphene and nanotubes~\cite{Guinea.2010, Homma.2009}, all parameters have been estimated from the 
experiments: The parameter $\alpha$ can be estimated from the charging energy $U\sim 10- 12 \meV$, and is found to be in the 
range $\alpha \approx  0.005 - 0.015\, \rm meV/nm^{2}$  (see Appendix~ \ref{app:charging_energy}). 
Here we shall use the value $\alpha =   0.015\, \rm meV/nm^{2}$ yielding the closest resemblance to the experimental data
of Ref.~\cite{Deshpande.2008}.
The radius $R= 1.6\rm \,nm$ is determined from the  
 curvature-induced gap $E_g\approx 220 {\rm \,meV}$ reported  in Ref.~\cite{Deshpande.2008} (and is directly related tho the effective mass, 
$E_g \approx 2m^*c^2$), 
and yields a spin-orbit splitting $\Delta_{\rm SO }\approx2.1\,\rm K$ (see Ref.~\cite{Kouwenhoven.2015}). 
Finally,  we have used the experimentally measured value,  $g_{\rm orb}\approx 5.8$~\cite{Deshpande.2008}.

Results of these simulations have been  summarized in Fig.~\ref{fig:phase_diagram_soi0}.
Though the magnetization pattern may be not as systematic as the ones reported in Ref.~\cite{Deshpande.2008}, 
-- possibly due to our approximate valence bond method, -- the similarity and the correct location of the 'phase boundaries' are, 
nevertheless, striking. The overall good agreement is, however, shaded by the fact that 
for these  parameters the density of the electron crystal starts to exceed the 
crossover value $n_e^*$ for $N\gtrapprox 26$  at the center (see Appendix \ref{app:charging_energy}). 
A somewhat weaker confinement, $\alpha \approx 0.01\, \rm meV/nm^{2} $ 
increases this characteristic value of $N$, and  yields also a charging energy in better agreement with the experimentally observed value, 
but the agreement of the 'phase diagram' in  Fig.~\ref{fig:phase_diagram_soi0} gets worse.

\section{Range of validity of the Wigner crystal description and scaling relations}
\label{sec:scaling}

Throughout our previous analysis, we assumed that electrons are reasonably localized
 by their strong Coulomb interaction. While this assumption is certainly not 
 correct for an infinite chain, where charge fluctuations are unlimited and 
 no long-ranged charge order exists even at $T=0$ temperature, it can certainly be applied in a finite system, where 
  charge fluctuations  are pinned. Then our approach is valid under the condition that 
 typical quantum fluctuations of the localized charges  be less than their separation,
  $ \Delta z \ll d$.  
  
  The ratio  $ \Delta z / d$ is directly related to the parameter $r_s$. 
  Computing the width of a Gaussian wave function 
 selfconsistently within the Coulomb  potential of an infinite chain of particles yields the simple
 estimate (see Appendix~\ref{app:ECP})
    \beq
  \frac{\Delta z}{d}  =  
   \left(\frac 1{{4 \,F \,r_s }}\right)^{1/4}\,
   \label{eq:extension}
  \eeq
with the geometrical factor $F$ depending on the  densities (wave functions) of the  other localized charges. 
For  perfectly localized particles we find  $F=\zeta(3)\approx 1.202$,  with $\zeta$ the Zeta function
(see Appendix~\ref{app:ECP}).  
However, the factor $F$ increases as one approaches the border of the Wigner crystal regime, $r_s\approx r_{1D}^*$, 
which we define  as the value of $r_s$, where the charge density  is suppressed by  a factor 
of 2 as one moves  from one lattice position to the next one, a condition  yielding 
 $ {d}/ \Delta z\approx  2.35$ for  Gaussian wave packets. Smearing the electron charges 
  then in boxes of width $2\,\Delta z$ yields $F\approx 1.70$ at the transition,  
 corresponding to the rough estimate, $r_s\approx r_{1D}^*\approx 4.4 $. 

A more accurate way to estimate $r_{1D}^*$ is to perform calculations for a molecule in a harmonic trap, 
where one can  squeeze the atoms together by increasing the confinement frequency 
$\omega_0$~\cite{Rontani.2010}. The crossover density and thus 
$r_{1D}^*$ can then  be determined by just looking at the separation $d$ of the two charges when the 
charge density at the center is reduced by a factor of $2$. This more accurate 
procedure yields the crossover value $r_{1D}^*\approx 3.3$, used throughout this paper.

Now we  show that at the crossover, $r_s=r_{1D}^*$, the exchange coupling $J^*$ and the density 
$n_e^*$ obey the scaling relations, Eq.~\eqref{eq:scaling}.  To prove this, we first observe that, according to our 
discussion in Sec.~\ref{sub:DerivationExchange}, the   electron-electron interaction 
 can be replaced by the angular averaged interaction, $U_0(z,R)$ in Eq.~\eqref{eq:h12-offdiag}.  
Introducing  the dimensionless coordinates, $\xi_i \equiv n_{e} z_i$, the Hamiltonian of 
the interacting particles becomes
\beq
 \label{eq:rescaled}
 H= \frac{\hbar^2 n_e^2}{m^*} \; {\cal H}\;,
 \eeq
 with the dimensionless Hamiltonian $ {\cal H}= {\cal H}(r_s,n_e R)$  given by 
 \beq
{\cal H}= \sum_{i}-\frac{1}{2}\,\frac{\partial^2}{\partial \xi_i^2}+ r_{s} \sum_{i<j}u_0(\xi_{ij},R \cdot n_e)\;,  
\eeq
where the dimensionless averaged Coulomb interaction $u_{0}$ trivially depends on the dimensionless parameter
$n_{e} R $.
Thus, in the dilute limit $n_e R \ll 1$ we have $ {\cal H}\approx  {\cal H}(r_s,0)$, and 
the structure of the dimensionless wave function and 
the energy spectrum of the dimensionless Hamiltonian $\cal H$ depend
only on $r_s$. It follows  immediately  that at the crossover point, $r_s  = r_{1D}^* \approx 3.3$, the density of the gas scales as

\beq
 n_e^* = \frac 1 {r_{1D}^*} \frac{e^2 m^*}{\epsilon\; \hbar^2 } \sim  {m^* \over \epsilon},   
 \label{eq:nestar}
\eeq
while  the exchange energy is just a  universal number ($A$) apart from the overall energy scale in 
Eq.~\eqref{eq:rescaled},
\beq
J^*= A \;  \frac{\hbar^2 {n_e^*}^2}{m^*} \sim \frac  {m^*}{\epsilon^2}\,.
 \label{eq:J_star}
\eeq
Eq.~\eqref{eq:J_star} just follows from the fact that, Ð in the spirit of the virial theorem, - at the crossover, the Coulomb, the kinetic, 
and the  exchange energies are all  approximately equal, while
Eq.~\eqref{eq:nestar}  states that  the density of the gas is inversely proportional to the effective Bohr radius.  

These general  scaling relations hold under the  condition $n_e^* R \ll 1$.  
Using the relation    $m^*= \hbar^2/(3R\gamma)$~\cite{Kouwenhoven.2015}
with   $\gamma = 0.54 \,\rm eV\cdot nm$  (yielding, e.g., 
$m^*\simeq 0.0294\; m_{e}$ for a nanotube of radius $R=1.6\,\rm nm$), 
this condition simplifies to
\beq
n_e^* R \approx \frac {0.27} \epsilon \ll 1\,.
\label{eq:WC_condition}
\eeq
This inequality  is well satisfied for even slightly screened nanotubes with $\epsilon \sim 2 - 3$, 
but we find that  relations \eqref{eq:scaling} are also 
obeyed by the exchange couplings and densities extracted from the two-body spectrum 
of unscreened nanotubes with $\epsilon =1$, for which \eqref{eq:WC_condition} is certainly 
only poorly satisfied.

\section{Summary and conclusions}

In this work, we attempted to account for the magnetic behavior of a Wigner crystal that forms in 
a confined semiconducting carbon nanotube.  We have carefully estimated   the exchange interaction ($J$)
between neighboring localized electrons (holes) in the crystalline state, and have shown that it 
is   SU(4) symmetric with  very good accuracy. 
For poorly screened small diameter semiconducting nanotubes, the microscopically  determined exchange couplings 
at the 'melting' of the crystal  turn out to be surprisingly    large ,  
 $J^*\sim 100 {\rm \,K}$. These large values 
follow from   very robust scaling arguments,  and are also  in agreement with 
experiments~\cite{Deshpande.2008,Ilani.2013} as well as independent   
theoretical computations~\cite{Rontani.2010,Ilani.2013}, also reproduced here.
As we argued, at the cross-over the exchange coupling $J^*$ is just proportional to the effective Bohr energy, 
with $m_e$ replaced by $m^*$ and with an additional factor $1/\epsilon^2$, yielding the simple estimate, 
 $J^*\sim m^*/(m_e \epsilon^2)\,\times 1\,{\rm Ry}$. Determining the numeric prefactor omitted here 
 requires more accurate  computations, but it is not unreasonable to assume it is in the range of $0.01- 0.1$. 
For a nanotube with $R\approx 1.6\,{\rm nm}$ and $\epsilon=2$, this heuristic estimate 
gives  $10\,\rm K$ to $100\,\rm K$, consistent with our more accurate calculations.

 Spin-orbit coupling ($\Delta_{\rm SO}$) breaks the SU(4) spin  symmetry down to 
 SU(2)$\times$SU(2)~\cite{StrunkPRB15}. 
As we demonstrated in Section~\ref{sec:WignerMolecules}, for small, $N=2$ -- $4$ particle Wigner molecules,
the interplay between  $J$ and $\Delta_{\rm SO}$ leads to  interesting spin excitation 
spectra  with excited states  classified as 
SU(2)$\times$SU(2) multiplets. 
This intriguing  spin spectrum should  readily be seen in the co-tunneling spectrum of Wigner  molecules,  
 would provide  direct information on $J$,  $\Delta_{\rm SO}$, and would also evidence the underlying SU(4) and the 
 residual SU(2)$\times$SU(2) symmetries. 

The interesting spin structure of the molecule can lead to exciting quantum states 
when the molecule is coupled to electrodes~\cite{ZarandGoldhaber2003,Kouwenhoven2005}. For small $\Delta_{\rm SO}$, the SU(4) spin, coupled 
to two (SU(4)) Luttinger liquids, e.g., may give rise to a SU(4) two-channel 
Kondo state, characterized by an anomalous scaling dimension $\Delta = 2/3$~\cite{LUDWIG1994}. Coupling the SU(4) Wigner 
molecule to side electrodes would, on the other hand, lead to an SU(4) Fermi liquid state,
while higher dimensional SU(4) spins may give rise to exotic underscreened Kondo states. 
A finite $\Delta_{\rm SO}$ will, however,  induce a crossover to SU(2)$\times$SU(2) states, 
and lead to less exciting  SU(2) Kondo physics at low temperatures. 

We remark that the competition between $J$ and $\Delta_{\rm SO}$ has even more exciting implications 
for homogeneous crystals~\cite{Moca_unpublished}. The spin-orbit coupling  breaks the original SU(4) 
spinon excitations of the SU(4) antiferromagnet  into SU(2) spinons 
propagating with 3 different spin velocities, and leads to a quantum 
phase transition with one of the spinons becoming gapped  as 
we move deeper into   the Wigner crystal regime.

To test our semi-microscopic approach, we have performed a detailed  modeling of the experiments of Ref.~\cite{Deshpande.2008}.  
We estimated the basic  model parameters ($m^*$, $g_{\rm orb}$ and confinement strength $\alpha$) 
 directly  from the experiments. We have set the unknown dielectric constant  to 
$\epsilon=2$~\cite{Homma.2009}.  Performing  a self-consistent valence-bond calculation
for an increasing number of electrons  in an external magnetic field, 
 we recovered the experimentally observed magnetic 'phase boundaries' with reasonable accuracy. 
 As we have discussed in Section~\ref{sec:inhom},  these 'phase boundaries' from the inhomogeneity of the crystal, 
 and do not correspond to new phases, rather they can be interpreted as 
 the emergence of different types  of antiferromagnetic domains  at the center  of the nanotube, where the density 
 and thus the exchange coupling are both the largest.
Given the 
 simple multiscale calculations we performed,  the good agreement with the experiments is  striking. 

 We should remark though that while the
 'phase boundaries' we get are qualitatively and quantitatively very  close to the experimentally 
observed ones, we do not observe regular  two-fold magnetization patterns, our patterns are closer to the 
ones presented in the supplemental information of Ref.~\cite{Deshpande.2008}.
This may be a consequence of the valence-bond approach we employed or possibly the 'melting' of the 
Wigner crystal for larger particle numbers, which for our parameters occurs at $N\gtrsim 26$.

We should also comment here on the value of the value of the dielectric constant $\epsilon$. 
It would be natural to assume that $\epsilon\approx 1$ in a nanotube suspended in vacuum~\cite{Guinea.2010,IlaniPrivate,Homma.2009}.
However such a small value of  $\epsilon$ is inconsistent with the data. For $\epsilon=1$, 
only a shallow parabolic confinement with $\alpha\lessapprox 0.003   \, \rm meV/nm^{2} $ can
 yield charging energies compatible with the experimentally reported values, 
$U\sim 10 \meV$ (see Appendix~\ref{app:charging_energy}). For such shallow confinement the unscreened Coulomb repulsion pushes 
the charges quickly towards the end of the nanotube,  and already for about $N\sim 10$ they form a homogeneous crystal 
all over the nanotube.  Such a homogeneous crystal is clearly incompatible with the experiments: in such a crystal 
exchange couplings are approximately equal, and a huge magnetization jump should occur at a critical magnetic field, 
$B\sim J/\mu_{\rm orb}$, not seen experimentally.  Furthermore, such a homogeneous Wigner crystal will not melt gradually, 
but would develop a sudden transition in the whole crystal once the melting condition $r_{s} \lesssim r_{1D}^{*}$ is satisfied. 

Thus the value  $\epsilon\approx 1$ seems to be incompatible with the experimental data. 
So are larger values of $\epsilon\gtrsim 4$.  For these large values a very large confinement would be needed to yield 
$U\sim 10 \meV$. By our scaling arguments, the exchange coupling should then be less than  $J^*\sim 10 \, {\rm K}$, clearly 
inconsistent with the high field phase boundary observed in Ref.~\cite{Deshpande.2008} 
and the corresponding exchange coupling, $J\sim 60\, {\rm K} $. Furthermore, in this case the crystal would melt very quickly, 
once a few particles enter the  tube. 

We thus conclude that only $\epsilon\approx 2 $ seems to give a consistent explanation for the data 
reported in Ref.~\cite{Deshpande.2008}. For this value of $\epsilon$ the first transition between the completely polarized state 
and the spin-antiferromagnet (see Fig~\ref{fig:phase_diagram_soi0}) occurs well in the Wigner crystal regime, however, for $N\sim 26$ the Wigner 
crystal should melt, and  our approach becomes  questionable. 
The description of this regime of a partially melted confined Wigner crystal and the crossover between the 
Wigner crystal and electron (Luttinger) liquid regimes is a true theoretical challenge. 

 While spin-orbit coupling  gives rise to interesting spin excitations in small molecules,  we find that 
 for a semiconducting nanotube of radius $R=1.6 {\rm \,nm}$, corresponding to the  gap measured in Ref.~\cite{Deshpande.2008}, 
 $\Delta_{\rm SO}$, does not have a large impact on 
 the magnetic states within the 'phase diagram'. 
 It eliminates the orbitally polarized phase at very small fields but, apart from that, 
 the 'phase diagram' remains  almost identical to that of an SU(4) symmetrical Wigner crystal  with $\Delta_{\rm SO}\to 0$.
 
 Let us finally comment on the general implications  of our results and their limitations.
Although we focused on semiconducting (zig-zag) nanotubes, 
 most of our considerations are very general, and also apply with trivial modifications  to metallic tubes with curvature or strain induced band gaps and semiconducting nanotubes of other chirality. In particular, the  scaling relations
Eq.~\eqref{eq:scaling} are very general, and 
imply that the range of applicability of the Wigner crystal 
picture  as well as the strength of the exchange coupling  depend extremely sensitively 
 on  microscopic parameters of the tube and 
details of the experimental setup; to observe the Wigner crystal and its 
 magnetic structure it is essential to avoid strong screening and to increase the effective mass of the particles as much as possible. In practical terms,  small radius or strongly strained suspended nanotubes of $\epsilon\lesssim 2$ are best to observe the detailed structure of the crystal.   Correspondingly, 
 while the nanotube studied in Ref.~\cite{Deshpande.2008},   
 is found to be in the Wigner crystal regime  for  electron numbers $N\lesssim 26$,   metallic nanotubes 
 with small strain-induced gap   laid on or close to a substrate are extremely unlikely to host Wigner molecules, and 
 should rather be described in terms of extended electron (hole) states.

\acknowledgements
We would like to thank Shahal Ilani, Christoph Strunk, Milena Grifoni, and especially to Andr\' as P\' alyi for important and fruitful discussions, 
and to R\u azvan Chirla for the careful reading of the manuscript. 
This work has been supported by the Hungarian research grant OTKA K105149, by NSF DMR Grant No. 1603243 (LG)
and by UEFISCDI Romanian Grant No. PN-II-RU-TE-2014-4-0432 (CPM).

\appendix

\section{Construction of the Wigner crystal state in a zig-zag carbon nanotube}\label{sec:hilbert-space}

\begin{figure}
\label{fig:BZ}
  \includegraphics[width=.4\textwidth]{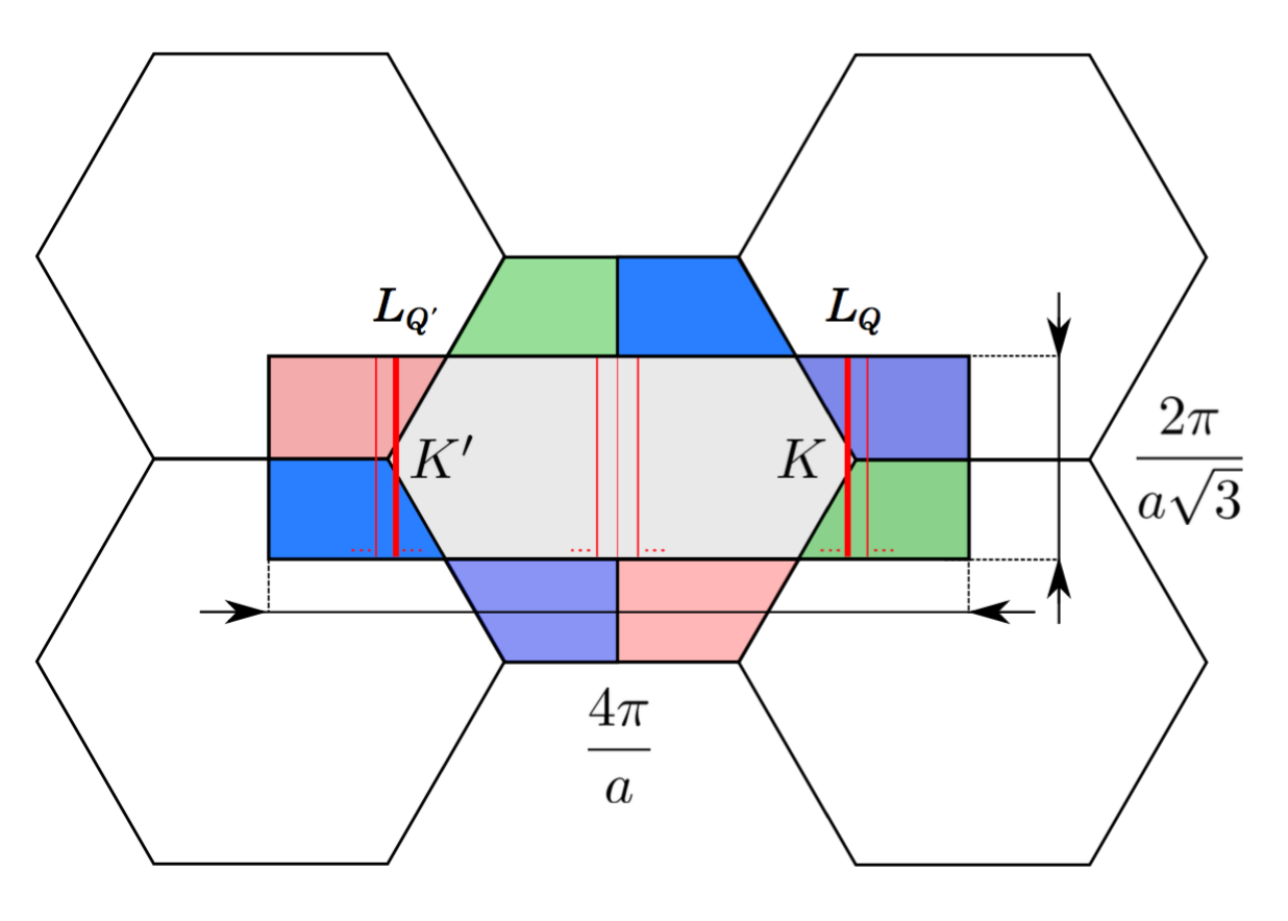}
  \caption{(Color online) Retailored Brillouin zone (BZ) of graphene (black rectangle).
The red lines denote states allowed for a zig-zag carbon nanotube, $K$ and $K'$ mark 
the two Dirac points,  $a$ is the lattice constant.  Thick solid red lines indicate the segments $L_{\pm Q}$.}
\end{figure}

In this Appendix  we  construct the microscopic wave function 
of the localized electrons forming the Wigner crystal. The Brillouin zone (BZ) of the 
underlying graphene sheet is represented in Fig.~\ref{fig:BZ}, with the two inequivalent
Dirac points  denoted as $K$ and $K'$. 
Rolling up the graphene  into a  carbon nanotube (CN) restricts the BZ to some parallel
line segments~\cite{Ando.2005}, with  their orientation  dictated by the chirality of rolling the nanotube. Here, for simplicity, 
we consider semiconducting  zig-zag CN's with chirality $(n,0)$ and radius  
$R=a\,n/2\pi$ ($n = 3 k + \nu$; $\nu=\pm1$). In this case,
 allowed states in  the graphene  BZ consist  of segments of length $2\pi/a\sqrt{3}$ parallel to the $k_{z}$ direction 
(vertical red lines in Fig.~\ref{fig:BZ}), and lowest lying excitations are on the segments closest to 
 $\bK=-\bK'= (4\pi/3a, 0)$. The minimum energy point of these vertical  segments is at the points
 ${\bf Q}=-{\bf Q}'= (Q,0)$ with $Q=(2n + \nu)/(3R)$.  For low density CNs it is enough to restrict ourself to these 
 two segments, $L_{\pm Q}=(\pm Q,q_z)$, indexed by the isospin quantum numbers $\tau = \pm1$.
Along these lines, for small $q_z$, excitations are 
 massive Dirac fermions with a dispersion 
\begin{equation}
  \varepsilon_\tau(q_z)\approx\pm\sqrt{c^2q_z^2+(m^*)^2c^4},
\end{equation}
with $c\approx 8\times 10^5 \,{\rm m/s}$ the Fermi velocity of graphene  and  $m^*$ the effective mass 
\begin{equation}
  m^*=\frac{\hbar}{3Rc}\;.
\end{equation}

The wave functions (of the unrolled nanotube) for the electrons  (holes)  can be expressed  by  Bloch's theorem as
$\psi^\pm_\mathbf{k} (\mathbf{r},\zeta) = \e^{\mathrm i\mathbf k\mathbf r} \, 
u_\mathbf{k}^\pm(\mathbf{r},\zeta)$. Here we explicitly separate the position 
vector into a two dimensional vector  within the graphene sheet, $\br$,  
and a coordinate $\zeta$ perpendicular to it. 
In cylindrical coordinates of the nanotube $\br=(R\varphi, z)$, 
and  the wave functions along the line segment $\bk \in L_{\tau Q}$ read
\begin{equation}
\label{eq:Bloch-fv}
 \psi^\pm_\mathbf{k} = \e^{\mathrm i Q  \tau \varphi}\, \e^{\mathrm i  q_z  z} \, u_\mathbf{k}^\pm(R\varphi, z, \zeta)
\end{equation}
with $r= R+\zeta$. These wave functions describe  particles of chirality $\tau $ circulating around the tube.

In the Wigner crystal, we create  wave packets from the  
states~\eqref{eq:Bloch-fv} 
\begin{equation}
\label{eq:wavepacket-def}
  \Psi^{\pm}_{j\tau}(\mathbf r,\zeta)= \int_{\mathbf k\in L_{\tau Q} } \dd q_z\, \, \e^{\mathrm i\mathbf k\mathbf r}\cdot u^{\pm}_{\mathbf k}(\mathbf r, \zeta)\cdot f (q_z)\;,
\end{equation}
with  $f (q_z)\propto \e^{-\mathrm{i} {\bar z} q_z}\e^{-\frac12 \Delta z^2 q_z^2}$  a Gaussian envelope,  
and  $\bar z$ the location of the wave packet  along the nanotube. 
Assuming that $u^{\pm}_{\mathbf k}$ only weakly depends  on $\bf k$, we 
obtain the quasiparticle wave function at position $\bar z = z_j^{(0)}$
\begin{equation}
\begin{split}
	 \label{eq:wc-hfgv}
		 \Psi^\pm_{j\tau\sigma} =\textstyle{\frac{1}{\sqrt{2\pi}}}\,&\e^{\mathrm i\, Q  \tau \varphi}
		 \textstyle{ \frac{1}{(\pi \,\Delta z)^{1/4}} }\e^{-\frac{(z-z_j^{(0)})^{2} }{2\,\Delta z^{2}}}\,\chi_{\sigma}
		\; u^\pm_{\tau {\bf Q}}(z,\,\varphi,\,\zeta),
\end{split}  
\end{equation}
with the $\pm$ sign referring to electrons and holes, and 
 $\chi_\sigma$ representing the spin part of the  wave function. The Bloch functions
$u_{\pm \mathbf Q}(\mathbf r, \zeta)$ in \eqref{eq:wavepacket-def} describe an almost homogenius 
background charge pattern, which varies only at the atomic scale, and can be ignored in many cases. 
%
We should emphasize that the single band approach presented here is only be valid for wide enough 
wave packets, 
\begin{equation}
\label{eq:crit.point}
\Delta z > \sqrt 3R.
\end{equation}

\section{The effective Coulomb potential}
\label{app:ECP}

\begin{figure}[!tb]
	\includegraphics[width=0.9\columnwidth]{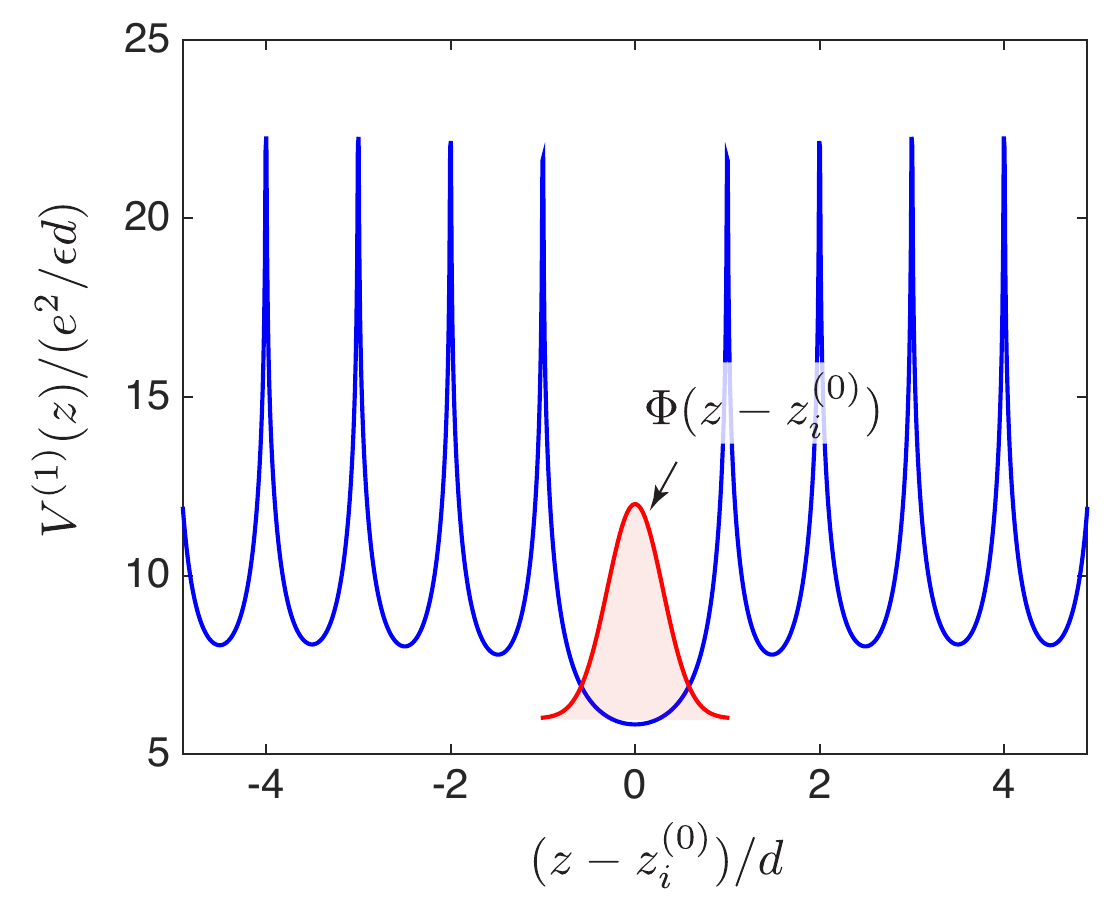}
	\caption{(Color online) The Hartree potential $V_{i}^{(1)}(z)$ confining the $i$'th electron, defined in Eq.~\eqref{eq:vi}. 
} 
\label{fig:vi}
\end{figure}
\label{sec:hartree_one-particle}
The Coulomb potential in cylindrical coordinates is 
\begin{equation}
\label{eq:u}
U(z,\varphi)=\frac{e^2}{\epsilon}\,\frac{1}{\sqrt{z^2+\bar \alpha^{2}+2R^2(1-\cos\,\varphi)}}
\end{equation}
with $e$ the electron charge and $\epsilon$ the relative dielectric constant. In the exact
diagonalization approach of Ref.~\cite{Rontani.2010} the microscopic cut-off   $\bar\alpha$
describes a crossover between 
the Coulomb potential and a Hubbard-like short range interaction for $z\to 0$, and  
was fixed to $\bar\alpha=e^{2}/U_{0}\,\epsilon$, with $U_{0}= 15\, {\rm eV}$.
Then, the average interaction felt by two electrons at a distance $z$ is
\begin{equation}
\begin{split}
\label{eq:ellip}
		 U_0(z)=\int_0^{2\pi}\,\frac{\dfi}{2\pi}\,U(z,\varphi).\\
		 =\frac{e^2}{\epsilon}\,{1\over |z|}\;f(z/R, \bar \alpha/R)
\end{split}  
\end{equation}
with $f(z/R, \bar\alpha/R)$ a dimensionless function
\begin{equation}\label{eq:f}
f\big({z\over R}, {\bar\alpha\over R}\big) = \frac{|z|}{\sqrt{z^{2}+\bar \alpha^{2}+4R^{2}}}K
\big( \frac{2R}{\sqrt{z^{2}+\bar \alpha^{2}+4R^{2}}}  \big)
\end{equation}
given in terms of  the complete elliptic integral of the first kind,  $K(x)$~\cite{Ryzhik}.
The screening length  $\bar \alpha$ in Eq.~\eqref{eq:f}  is of the order $\bar \alpha\sim 0.1 \,{\rm nm}/\epsilon$, 
is much smaller than $R$, and regularizes the potential in the limit $z\to 0$, while 
for large distances, $z\gg R $ the usual Coulomb behavior is recovered,
\begin{equation}\label{eq:U_free}
 U_0(z\gg R)\approx \frac{e^2}{\epsilon}\, \frac{1}{|z|}.
\end{equation}
The Hartree  potential felt by particle $i$   is  given by $V_i^{\text{(1)}}(z) = V^{\text{(1)}}(z-z_i^{(0)})$
 with
\begin{equation}
\label{vi-kieg}
 V^{\text{(1)}}(z)=\sum_{j\neq 0} \int \dd z' \frac{\dd \varphi'}{2\pi}\frac{\dd \varphi}{2\pi} U(z-z',\varphi-\varphi')| \Psi_{j\tau\sigma}(z',\varphi',\zeta)|^2.
\end{equation}
Deep in the Wigner crystal the wave functions are well localized, 
and to a good approximation 
\begin{equation}
\label{eq:vi}
 V^{(1)}(z) \approx  \sum_{j\neq 0} U_0(z-z_j^{(0)}), 
\end{equation}
with $U_0(z)$ given by Eq.~\eqref{eq:ellip}.  The resulting Hartree potential is shown in 
Fig.~\ref{fig:vi}. A similar procedure yields the  two-particle potential displayed in Fig.~\ref{fig:vij}.

We now estimate the extension of the wave function in this Hartree potential. 
We first   approximate  $U_0(z-z_j^{(0)})$ by a Coulomb potential to  
obtain the following parabolic approximation by expanding 
\eqref{eq:vi}, 
$$
 V^{(1)}(z) \approx  V^{(1)}(0)+  \frac{2 e^2}{\epsilon \,d^3} \zeta(3)\, z^2\;,
 $$
with $d=n_e^{-1}$ the separation of electrons and $\zeta(3)=\sum_{k=1}^\infty 1/k^3\approx 1.202$~\footnote{For an infinite chain we 
have $ V^{(1)}(0)\to \infty$, a divergence compensated by the background of positive charges.}. 
Solving the harmonic oscillator problem in this harmonic potential yields then 
the simple estimate, Eq.~\eqref{eq:extension}.

\section{Exchange interaction in a Wigner molecule}\label{app:WM}
\begin{figure}[!tb]
\includegraphics[width=0.9\columnwidth]{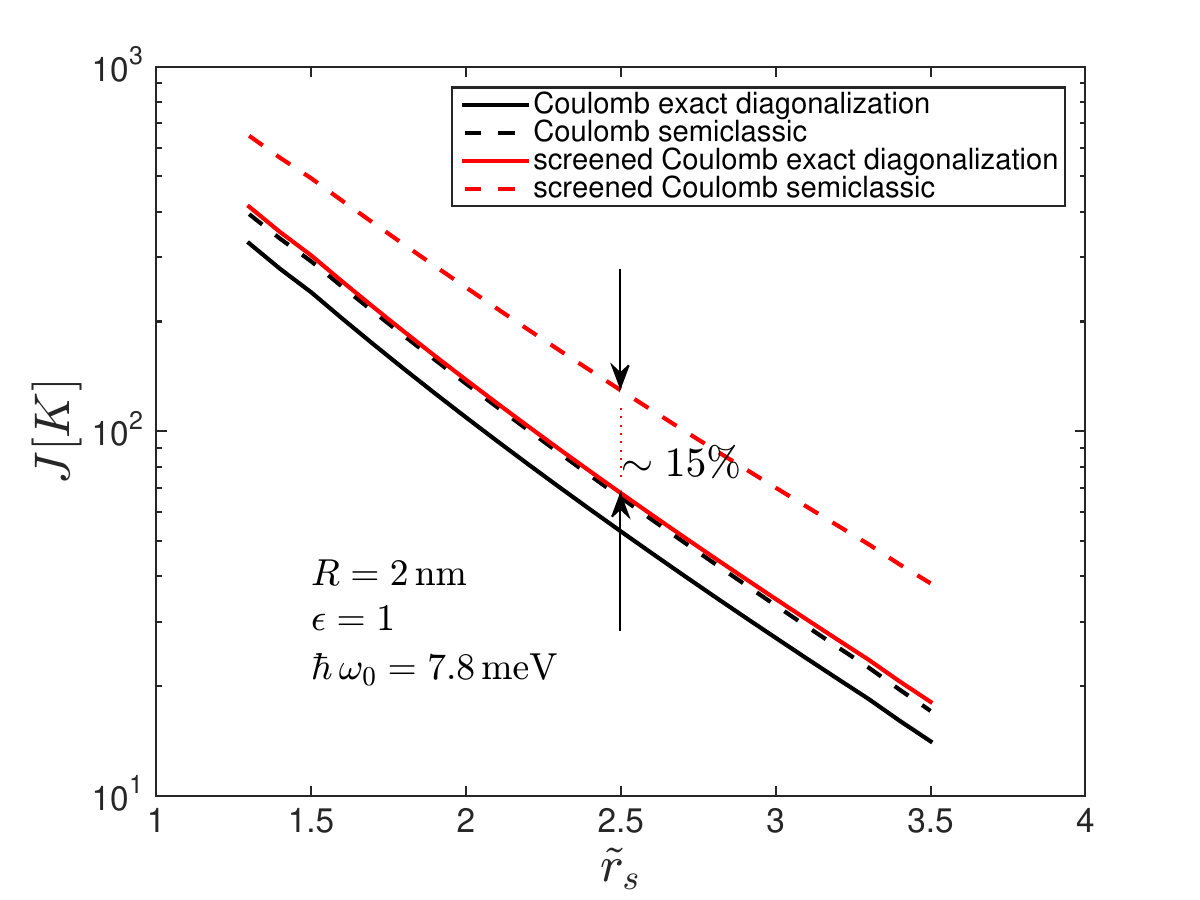}
\caption{The exchange interaction as function of the dimensionless ratio $\tr_{s}$
for a Wigner molecule. The radius of the nanotube was fixed to $R=2\, \rm nm$ and the 
confinement energy is $\hbar\,\omega_{0}= 7.8\rm\, meV$. The solid lines
are obtained using the pure Coulomb interaction~\eqref{eq:U_free} while 
for the dashed lines the effective potential in Eq.~\eqref{eq:ellip} with
$\bar \alpha=0$ has been used.
}  
\label{fig:J_comparison}
\end{figure}
In this appendix  we present the semiclassical approach to determine the 
exchange interaction $J$, which we 
also  compare with the results of  exact diagonalization~\cite{Rontani.2010,Ilani.2013}.   
We consider two interacting electrons of mass $m^*$ in a parabolic confining potential  of frequency 
$\omega_0$. The Schr\" odinger equation can be  factorized in this case in terms of the relative ($z=z_2-z_1$) and center of 
mass ($Z = (z_1 + z_2)/2$) coordinates, $H=H_{\rm rel}(z) + H_{\rm COM}(Z)$. The center of mass
motion is that of a  harmonic oscillator of frequency $\omega_0$, and is completely decoupled   
from  the relative motion described by the single particle Hamiltonian,  
\begin{equation}
H=-\frac{\hbar^{2}}{2 \mu} \frac{\partial^{2}}{\partial z^{2}}+ {1\over 2}\, \mu\, \omega_{0}^{2}z^{2}+
\frac{e^{2}}{\epsilon}\frac{1}{|z|}f\big({z\over R}, {\alpha\over R}\big)\label{eq:H2},
\end{equation}
with  $\mu=m^{*}/2$  the reduced mass and 
$f(z/R,\bar \alpha/R)$ the cut-off function in Eq.~\eqref{eq:f}. 
With a good accuracy, we can set the parameter $\bar \alpha/R$ to zero.  
We can then  make the Hamiltonian dimensionless by 
introducing the dimensionless coordinate, $\rho =z/\lambda$, with 
$\lambda=(\hbar/m^{*}\omega_{0})^{{1/2}} $ the non-interacting oscillator  length, 
and dividing it by the  natural energy scale, $ \hbar \omega_{0}$. 
In these units, the Hamiltonian becomes
\begin{equation}
{\cal H}={\cal T}+{\cal V}(\rho)=-\frac{\partial^{2}}{\partial \rho^{2}}+ {1\over 4} \rho^{2}+
\widetilde{r}_{s}\frac{1}{|\rho|}f\big(\rho \,{\lambda\, \over R}\big)\;,\label{eq:H2d}
\end{equation}
with 
$\widetilde{r}_{s}$ characterizing the strength of Coulomb interaction compared to that of the parabolic 
confinement,
\begin{equation}\label{eq:tr_s}
\widetilde{r}_{s} = \frac{e^{2} }{\hbar \omega_{0}\,\epsilon\lambda} .  
\end{equation}
Notice that   $\widetilde{r}_{s}$  in Eq.~\eqref{eq:tr_s} is different from the usual  $r_{s}$,
defined by Eq.~\eqref{eq:r_s}.

The dimensionless potential  $ {\cal V}(\rho)$ in Eq.~\eqref{eq:H2d} displays two minima at $\pm \rho_{0}$, corresponding to the  ground state positions of the classical particles. Close to these minima, the potential can be approximated by 
parabolas, and the molecule vibrates with a frequency
$\Omega = \omega_0 (2 {\cal V}^{\prime\prime} (\rho_0))^{1/2}$, 
where ${\cal V}^{\prime\prime}(z)$ is the second derivative with respect to $z$.
 Tunneling processes between $\pm\rho_0$
give rise to a splitting of these two levels, which we can identify  as the  exchange coupling. 
At the semiclassical level, we can thus estimate $J$ as the tunneling amplitude~\cite{Matveev.2004}  
\begin{equation}
\label{eq:J}
J\approx \frac{\hbar\,\Omega}{\pi}\,\e^{- \int_{-A}^A\dd \rho\,\sqrt{{\cal V}(\rho)-
({\Omega}/{2 \omega_0})}}, 
\end{equation}
with $A$ denoting  the classical turning-point determined by the equation 
${\cal V}(A) =\Omega/2\omega_0$.  Alternatively, we can  determine the spectrum of 
Eq.~\eqref{eq:H2d} numerically, and extract the ground state splitting from there. 

Fig.~\ref{fig:J_comparison}
displays a comparison of the results of these two approaches as  a function of $\widetilde r_{s}$
for a nanotube of radius $R=2 \,{\rm nm}$ in a confining potential of frequency  $\hbar \omega_0 = 7.8 \,{\rm meV}$. 
Both  approaches  yield  an exponential decay of $J$ with increasing $\widetilde r_{s}$. 
The semiclassical method slightly overestimates the exchange coupling, but it gives a surprisingly accurate 
estimate for $J$. For a simple Coulomb 
interaction between the two electrons, \eqref{eq:U_free}, it estimates  
  $J$'s within $\sim 5 \%$, but its accuracy remains around$~\sim 15\,\%$ 
 for the more appropriate nanotube interaction,  Eq.~\eqref{eq:ellip}, too.
\begin{figure}[t]
\includegraphics[width=0.95\columnwidth]{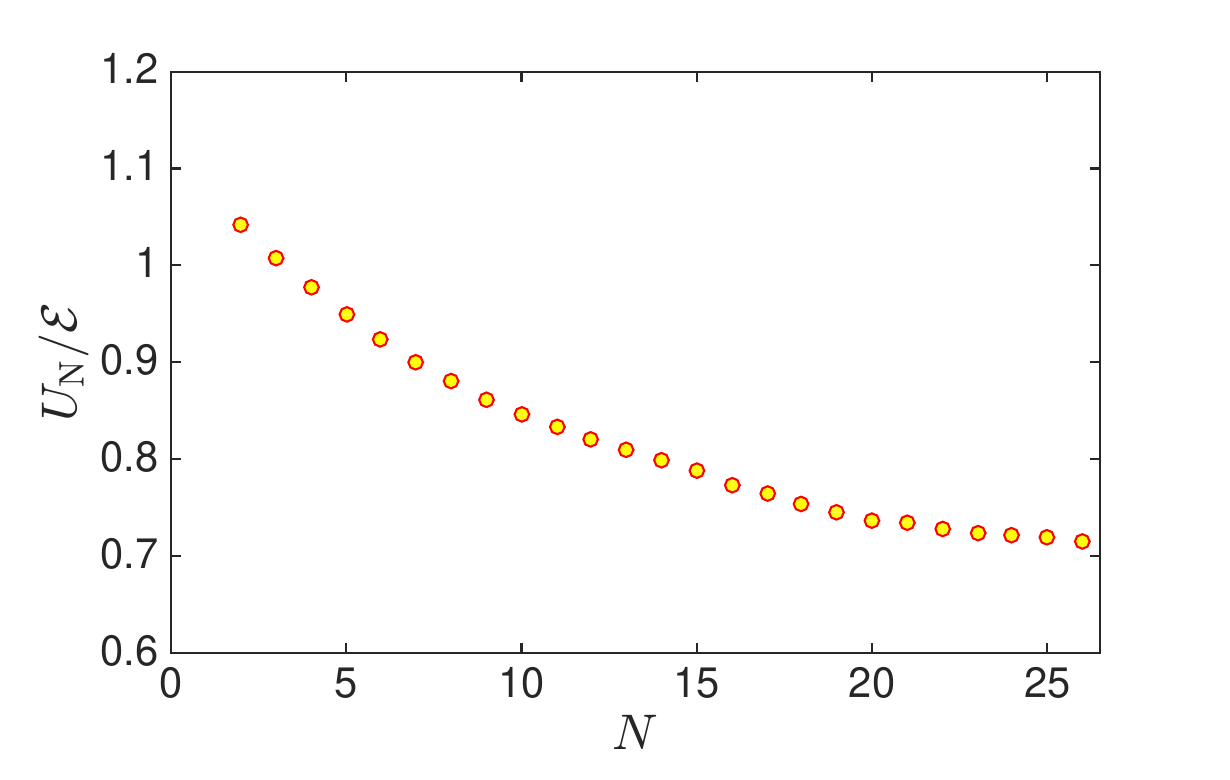}
\caption{Dimensionless  charging energy as function of the particle number $N$.}  
\label{fig:U_charge}
\end{figure}

%

\section{Charging energy of carbon nanotube in a confining potential}
\label{app:charging_energy}

The Coulomb energy of a nanoscale object  often depends approximately quadratically on the number 
of charged particles, 
\beq
E_{\rm Coulomb} \approx \frac U 2 N (N-1)\;.
\label{eq:ECoulomb}
\eeq
The charging energy $U$ can be directly extracted from  the Coulomb diamonds.
The data presented in   Ref.~\cite{Deshpande.2008}, e.g.,  yield a value  
$U\simeq 10-12\,\rm meV$. 
\begin{figure}[t]
\includegraphics[width=0.95\columnwidth]{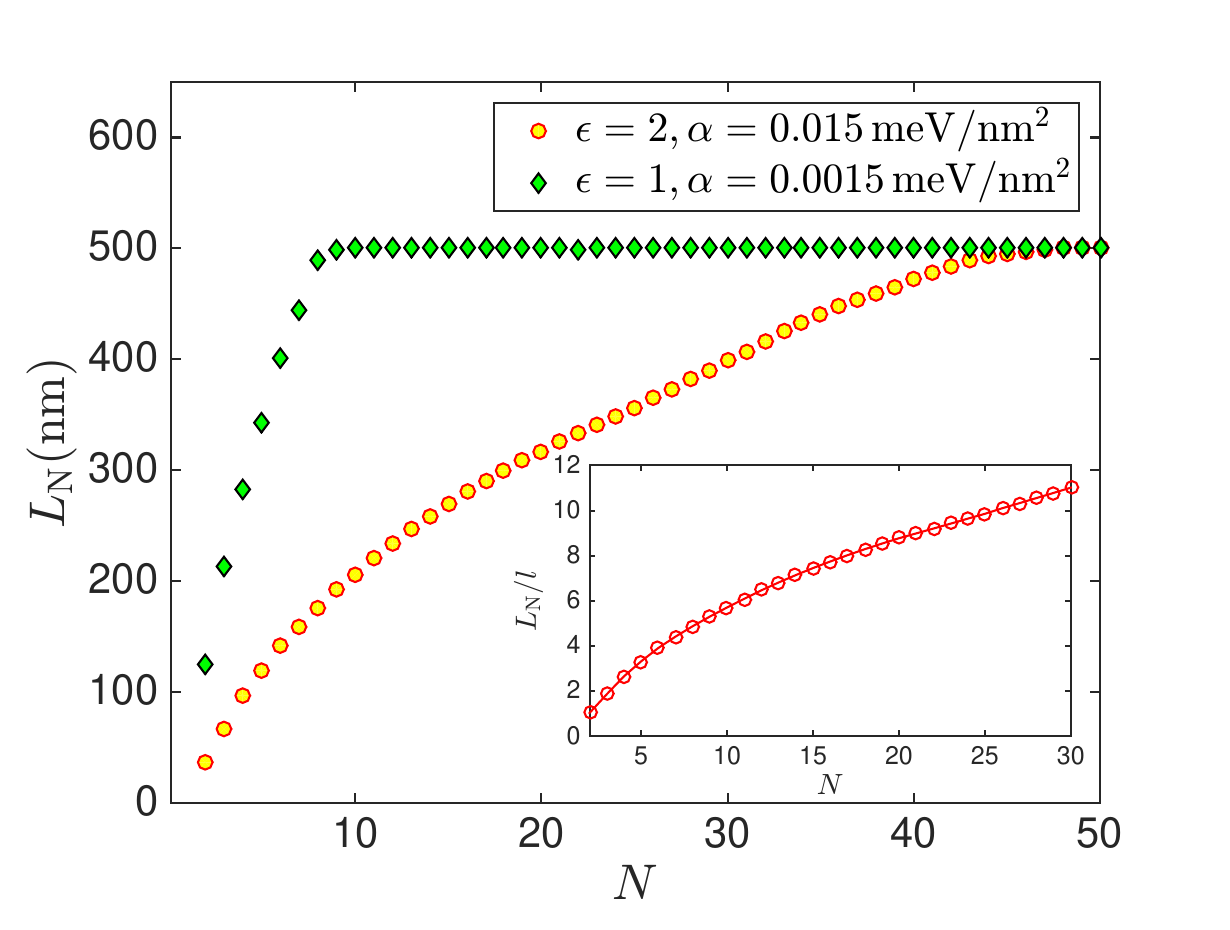}
\caption{ The 'length' of the Wigner crystal $L_{N}$, as a function of the number of charges $N$, for various potential depths $\alpha$ and 
dielectric constants. The length of the nanotube is $L=500\,\rm  nm$. 
Inset: $L_N/l$ for an infinite nanotube with parabolic confinement.
$L_N/l$ is a universal function of $N$.}.  
\label{fig:L_N}
\end{figure}

In this appendix we determine the effective value of $U$ as a 
function of the particle number $N$ for  a CNT confined by a harmonic potential. 
Starting from the ansatz \eqref{eq:ECoulomb},  the value of 
$U$ can be identified as the difference in the energy needed to add two consecutive electrons~\cite{Kouwenhoven.2006}, 
\begin{equation}
U \approx U_{N}\equiv \Delta E_{N+1}-\Delta E_{N},
\end{equation}
with $\Delta E_{N}= E_{N+1}-E_{N}$, and $E_{N}$ 
the total energy of the CNT with $N$  confined charges.

In a parabolic confinement, the energy $E(\{ z_i\})$ of a  given classical
charge configuration is the sum of the harmonic potential and the Coulomb energy, 
\begin{equation}
E(\{ z_i\})=\sum_{i=1}^{N}\frac{\alpha}{2}z_{i}^{2}+\sum_{i<j}\frac{e^{2}}{\epsilon|z_{i}-z_{j}|}.
\label{eq:E_N}
\end{equation}
For each $N$,  we first determine the coordinates $z_{i}$ of the particles by minimizing 
Eq.~\eqref{eq:E_N},  and then compute the total energy $E_{N}$. Introducing the dimensionless 
coordinates $\zeta_{i}= z_{i}/l$, with 
$l = (e^{2}/\alpha\epsilon)^{1/3}$, the potential energy becomes
\begin{equation}
E_{N} = {\cal E}\Big( \frac{1}{2}\,\sum_{i=1}^{N}\zeta_{i}^{2}+\sum_{i<j}\frac{1}{|\zeta_{i}-\zeta_{j}|}  \Big )\,.
\end{equation}
with the characteristic energy scale 
$$
{\cal E } = (e^{4} \alpha/\epsilon^{2})^{1/3}.
$$ 
As a consequence, $U_N$ can be expressed as 
$
U_{N} = {\cal E}\, f(N)\,.
$
We determined the universal function $f(N)$ numerically, and displayed it in  Fig.~\ref{fig:U_charge}.   
It has a weak dependence on $N$, and for  particle numbers of interest  $f(N) \approx 0.75$, yielding the relation
$$
U^{\rm eff}\approx 0.75\times {\cal E}.
$$  
This equation allows us to relate the screening parameter $\epsilon$ and the confinement parameter 
$\alpha$ through the experimentally determined charging energy. For $\epsilon = 2$, used throughout this work, $\alpha =  0.015 \,\rm meV/nm^{2}$ yield $U^{\rm eff}\approx 14.7\, {\rm meV}$, 
roughly consistent with the data. 
For $\epsilon =1$, however, one needs to use a much shallower confining potential
with $\alpha \approx 0.0015\,\rm meV/nm^{2}$ in order to be consistent with the experimentally observed 
charging energy. 
\begin{figure}[tb]
\includegraphics[width=0.95\columnwidth]{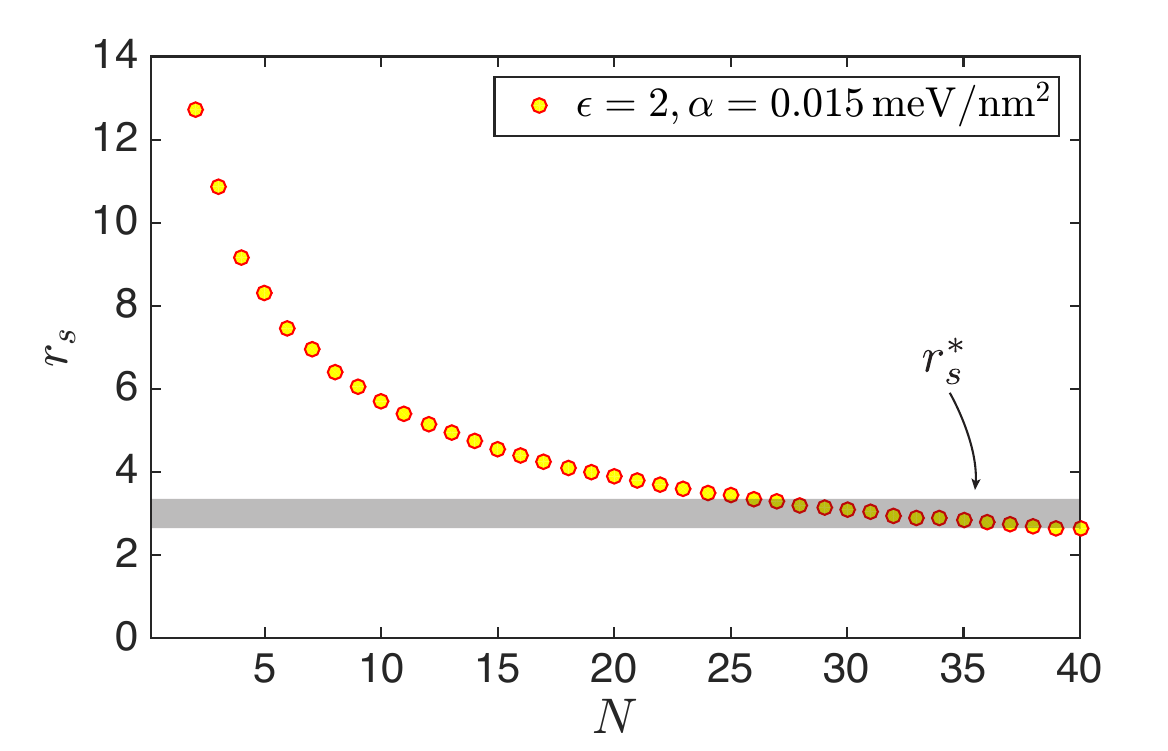}
\caption{ The dimensionless parameter $r_{s}$ as a function of the number of charges $N$
in the middle of a $500\, \rm nm$ long nanotube for $\alpha = 0.015\, \rm meV/nm^{2}$ and 
$\epsilon=2$. The Wigner crystal starts to melt at the center for $N=26$, once $r_{s}$
decreases below the crossover value $r_{s}^{*}\approx 3.3$.}
\label{fig:rs_N}
\end{figure}

In a finite nanotube of size $L$, the 'length' of the Wigner crystal $L_{N}$ 
increases monotonously with $N$ up to 
the  size of the tube. In an infinite nanotube, 
the growth is universal and can be expressed as $L_{N}/l = g(N)$, with $g(N)$
an universal function that we determined numerically. It is represented in  the inset of Fig.~\ref{fig:L_N}. In the main panel of the same figure, we represent $L_{N}$  as function of the number of charges $N$,  for  $\epsilon =2$ and $\alpha \approx 0.015\,\rm meV/nm^{2}$
as well as for $\epsilon =1$ and $\alpha \approx 0.0015\,\rm meV/nm^{2}$. 
While in the former case we can place 
about $N_{\rm max} \sim 40 $ electrons on the nanotube before hitting the walls, 
in the 'unscreened' case, $\epsilon = 1$,  this number is only $N_{\rm max} \sim 8 $. 
Beyond this number the separation of the particles becomes quickly equidistant, yielding an almost uniform 
exchange coupling. 

It is therefore evident that for large confinement potential 
depth $\alpha$ and strong screening $\epsilon$ a 
far larger number of charges can be squeezed inside the nanotube, as 
$l\propto (\alpha\,\epsilon)^{-1/3}$ decreases, but that's not a guarantee that the 
Wigner crystal state survive as $N$ increases. The only relevant quantity that controls
the 'melting point' of the Wigner crystal is the dimensionless parameter $r_{s}$. As displayed in 
inset of Fig.~\ref{fig:dij,jij}, for a given configuration with $N$ charges in the tube, $r_{s}$ is site 
dependent and has the smallest value in the middle of the chain. In Fig.~\ref{fig:rs_N} we represent
$r_{s}$ in the middle of a $500\, \rm nm$ long nanotube as function of the number of charges $N$.
When only a few charges are confined to the nanotube, 
the gas is diluted and, as expected, $r_{s}\gg r_{1D}^{*}$, 
but as the charges accumulate, $r_{s}$ decreases monotonously, and at some critical 
occupation $N_{\rm crit}\approx 26$ it reaches $r_{1D}^{*}$, and the crystal starts to melt in the middle. Adding more charges, 
$r_{s}$ decreases further, and the melting progresses towards the sides of the chain.

\bibliography{references}

\begin{thebibliography}{57}
\bibitem [{Ila()}]{IlaniPrivate}%
  \BibitemOpen
  \href@noop {} {}\bibinfo {note} {In suspended nanotubes charge distributions
  seem to be consistent with $\epsilon\approx 1$. (S. Ilani, private
  communication.)}\BibitemShut {NoStop}%
\end{thebibliography}

\end{document}